\documentclass{article}
\usepackage{latexsym}
\usepackage{amsmath,amsthm}
\usepackage{amssymb}
\usepackage{graphics}
\usepackage{color}
\newtheorem{remark}{Remark}[section]
\newtheorem{lemma}{Lemma}[section]
\newtheorem{theorem}{Theorem}[section]
\newtheorem{proposition}{Proposition}[section]

\def\Sb{\underline{S}}
\def\pa{\partial}

\newcommand{\nabb}{\mbox{$\nabla \mkern-13mu /$\,}}
\begin{document}
\title{The red-shift effect and radiation decay
on black hole spacetimes}
\author{Mihalis Dafermos\thanks{University of Cambridge,
Department of Pure Mathematics and Mathematical Statistics,
Wilberforce Road, Cambridge CB3 0WB United Kingdom}
\and
 Igor Rodnianski\thanks{Princeton University,
Department of Mathematics, Fine Hall, Washington Road,
Princeton, NJ 08544 United States}
}
\maketitle
\begin{abstract}We consider solutions to the 
linear wave equation $\Box_g\phi=0$ on a 
(maximally extended) Schwarzschild spacetime.   
We assume only that the solution decays suitably at spatial infinity on a complete
Cauchy hypersurface
$\Sigma$. 
(In particular, the support of $\phi$ may contain the bifurcate
event horizon.)
It is shown that the energy flux of the solution
through arbitrary achronal subsets of the black hole exterior region is bounded by
$C(v_+^{-2}+u_+^{-2})$, where $v$ and $u$ denote the infimum of
the 
Eddington-Finkelstein advanced and retarded time
of the subsets, and $v_+$ denotes $\max\{1,v\}$, etc.
(This applies in particular to subsets of the event horizon or null infinity.)
It is also shown that $\phi$ satisfies the pointwise decay 
estimate $|\phi|\le Cv_+^{-1}$ in the entire exterior region,
and the estimate $|r\phi|\le C_{\hat{R}}(1+|u|)^{-\frac12}$ in
the region $\{r\ge \hat{R}\}\cap J^+(\Sigma)$, for any $\hat{R}>2M$.
The estimates near the event horizon exploit an integral energy identity
normalized to local observers. This estimate can be thought
to quantify the celebrated red-shift effect.
The results in particular give an independent proof
of the classical result $|\phi|\le C$
of Kay and Wald without
recourse to the discrete isometries of spacetime.
\end{abstract}
\section{Introduction}
The concept of a black hole is a central one in general relativity:
Spacetime is said to contain
a black hole when it admits a complete null infinity
whose past has a regular future boundary. This 
boundary is called the event horizon
and the black hole itself is defined to be its future; 
the past of null infinity is known as the black hole exterior.

The simplest solutions of the Einstein vacuum equations of general 
relativity,
\begin{equation}
\label{vac}
R_{\mu\nu}=0
\end{equation}
containing black holes, 
the one-parameter Schwarzschild family of solutions, 
were written down in local coordinates~\cite{Schwp} 
in 1916, but only 
correctly understood as
describing spacetimes with black holes in the sense above, around 1960. 
Until that time, there were many arguments in the physics 
literature (e.g.~\cite{Einstein}) purporting to show that 
such solutions would be pathological and unstable. 
It was only when Kruskal demonstrated~\cite{kruskal}
that the event horizon could
be covered by regular coordinates that its true geometric character became clear,
and the problem of stability could be given a sensible and well-defined formulation. 

The Schwarzschild family turns out to be a sub-family of the two-parameter
Kerr family which describe stationary rotating black holes.
In its proper rigorous formulation,
the problem of nonlinear stability of the Kerr family is one of
the major open problems in general relativity.\footnote{In particular, 
it is conjectured that perturbations of
Schwarzschild initial data should 
evolve into a spacetime with complete null infinity whose past
``suitably'' approaches a nearby Kerr exterior.} At the heuristic level, however,
considerable progress has been made in the last 40 years
towards an understanding of the issues involved.
In particular, a
very influential role was played by the work of R.~Price~\cite{rpr:ns} in 1972, who
discovered a heuristic mechanism allowing for the decay
of scalar field linear perturbations on the Schwarzschild exterior. 
The mechanism depends on the following fact, known as the \emph{red-shift effect}, 
which had been understood previously 
in the context of geometric optics.
Given two observers $A$ and $B$, depicted below\footnote{What is depicted 
is the Penrose
diagram of a subset of Schwarzschild. See Section~\ref{Schwarzschild}.
The reader unfamiliar with these diagrams
can refer to~\cite{dr1}.},
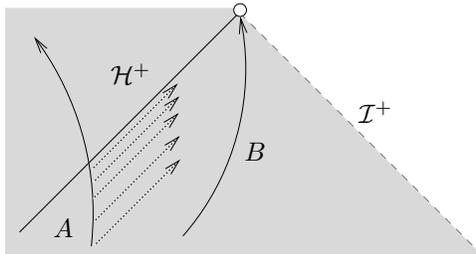
\[
\input{redshift.pstex_t}
\]
then if $A$ emits a signal at a constant rate with respect to his own proper time,
the frequency of the signal as received by $B$ is infinitely shifted to the red
as $B$'s proper time goes to infinity.

For spherically symmetric solutions of the coupled Einstein-scalar field system, and more
generally, the Einstein-Maxwell-scalar field system, the heuristic picture put forth
by Price is now a theorem~\cite{dr1}, and the red-shift effect described above
plays a central role in the proof. Moreover, one of the results of~\cite{dr1},
namely the decay rate 
\begin{equation}
\label{EHdecay}
|\phi|+|\partial_v\phi|\le Cv_+^{-3+\epsilon}
\end{equation}
along the event-horizon\footnote{Here, $v$ is a naturally defined 
Eddington-Finkelstein-like advanced time
coordinate.},
has important implications for the causal structure of the interior of the black hole:
In~\cite{cpam}, it is shown using $(\ref{EHdecay})$ that, in the charged
case, the spacetimes of~\cite{dr1}
do not generically terminate in everywhere spacelike singularities as 
originally widely thought, 
but rather, their future boundary has a null component across
which the spacetime metric can be continuously extended.
In particular, this implies that the $C^0$-inextendibility
formulation~\cite{chr:givp}  of
strong cosmic censorship is false for the system considered.

For the Einstein vacuum equations $(\ref{vac})$ in the absence of 
symmetry assumptions, 
results analogous to~\cite{dr1, cpam} seem out of reach at present. Clearly,
a first step towards attacking geometric non-linear stability questions is proper
understanding of the linear theory in an appropriate geometric setting.
This will be the subject of the present paper.
Our main result is the following:
\begin{theorem}
\label{main}
Let $\phi$ be a sufficiently regular 
solution of the wave equation
\begin{equation}
\label{Lwe}
\Box_g\phi=0
\end{equation}
 on the (maximally extended)
Schwarzschild spacetime $(\mathcal{M},g)$, 
decaying suitably at spatial infinity on an arbitrary 
complete asymptotically flat Cauchy surface $\Sigma$.
Fix retarded and advanced Eddington-Finkelstein coordinates $u$ and $v$
on one of the exterior regions. For any
achronal hypersurface $\mathcal{S}$ in the closure of this region, 
let $F(\mathcal{S})$ denote the
flux of the energy through $\mathcal{S}$, where energy is here measured with
respect to the timelike Killing vector field. Let 
$v_+=\max\{v, 1\}$, $u_+=\max\{u, 1\}$, and
$v_+(\mathcal{S})=\max\{\inf_{\mathcal{S}} v,1\}$,
$u_+(\mathcal{S})=\max\{\inf_{\mathcal{S}} u,1\}$. We have
\begin{equation}
\label{fluxdecay}
F(\mathcal{S})\le C((v_+(\mathcal{S}))^{-2}+(u_+(\mathcal{S}))^{-2}).
\end{equation}
(We also allow $\mathcal{S}$ to be a subset of null infinity, interpreted
in the obvious limiting sense.)
In addition, we have the pointwise decay rates
\begin{eqnarray}
\label{pointwise}
\nonumber
|\phi|&\le& Cv_+^{-1} \quad {\rm\ in\ }
\quad \overline{J^-(\mathcal{I}^+)
\cap J^+(\mathcal{I}^-)}\\  
|r\phi|&\le& 
C_{\hat{R}}(1+|u|)^{-\frac12} \quad {\rm\ in\ }  \quad
\{ r\ge \hat{R}>2M\}\cap J^+(\Sigma).
\end{eqnarray}
\end{theorem}
In the spherically symmetric case, the above result follows from
a very special case of~\cite{dr1}. (See also~\cite{st:pc}.) Decay for $\phi$, without
however a rate, was first proven in the thesis of Twainy~\cite{ft:td}.
The uniform boundedness of $\phi$ is a classical result of Kay and Wald~\cite{kw:lss}.
For the more general
Kerr family, even uniform boundedness remains an open problem 
(see however~\cite{fksy}).

We should also note that, independently of us, a variant of the problem
considered here is being studied  by  \cite{morebs}.

A statement of Theorem~\ref{main} in local coordinates will be given later.
This, in particular, will explain the dependence on $C$, and the minimum
regularity assumptions necessary. The reader wishing to penetrate 
deeper into this problem, however,  is strongly encouraged to learn 
the language necessary for the above geometric formulation. For neither
the correct conditions on initial data, nor the desired 
statement of decay, have particularly natural formulations when 
stated with respect to Regge-Wheeler coordinates.\footnote{In particular,
it is not correct to restrict to ``compactly'' supported initial data in
Regge-Wheeler coordinates, because $r^*=-\infty$ corresponds to points
in the actual spacetime. These issues are well known in the 
relativity community, and the reader should consult the nice
discussion in~\cite{kw:lss}.} The motivation
for considering decay as stated in $(\ref{fluxdecay})$ is that
it is this formulation that has direct relevance both
in the astrophysical regime, as well as for the fate
of observers who enter the black hole
region.
For $(\ref{fluxdecay})$ applied to $\mathcal{S}=\mathcal{I}^+$
gives decay rates for the energy radiated to infinity (this is what is astrophysically
observable), while applied to $\mathcal{S}=\mathcal{H}^+$,
it gives decay rates for the energy thrown into the black hole (this is what concerns
the observer entering the black hole).

It is interesting to note that
the decay rates $(\ref{fluxdecay})$--$(\ref{pointwise})$ are
sufficiently fast so as to suggest that the picture established in~\cite{cpam}
may remain valid in the absense of symmetry
assumptions, in particular, the existence of a marginally trapped tube
which becomes achronal and terminates at $i^+$, and a weak null singular boundary
component\footnote{For even the weakest results of~\cite{cpam}, one
still requires 
the analogue of $\phi\le Cv^{-\frac12-\epsilon}$ along the event
horizon,
for an $\epsilon>0$.} to spacetime, 
emanating from $i^+$, across which the spacetime
can still be continuously extended.

The techniques of this paper are guided by the principle that they
should be relevant for non-linear stability 
problems. For such problems, in the absence of symmetry, 
energy-type estimates have proven the most
robust~\cite{book}. 
For Lagrangian theories like the homogeneous wave
equation, such estimates naturally arise 
by contracting suitable vector fields $V^\alpha$ with the energy-momentum tensor 
$T_{\alpha\beta}$, to produce a one form $P_\alpha$. (See~\cite{book2} for
a general discussion.) The divergence
theorem relates the spacetime integral of $\nabla^\alpha P_\alpha$
with suitable boundary terms. The method can be used 
to estimate the spacetime integral from the boundary
terms, but also the future boundary terms from the past boundary and the spacetime integral. 
As we shall see below, both implications
will be used here.

Let us recall the situation for the wave equation $(\ref{Lwe})$ on
Minkowski space. The technique described here was introduced by
Morawetz, see \cite{mo} and Klainerman~\cite{kl}. By applying the 
method of the previous paragraph
 to the Killing vector field $\frac{\partial}{\partial t}$, one obtains
the usual energy conservation. By applying
the method to the conformally Killing ``Morawetz'' vector field 
\[
K=v^2\frac{\partial}{\partial v}+u^2\frac{\partial}{\partial u},
\]
in the region $\{1\le t\le t_i\}$, one obtains an identity relating
a spacetime integral in this region to boundary integrals
of weighted energy densities on $\{t=1\}$ and $\{t=t_i\}$.
The spacetime integral can be completely
removed by a second application of the
divergence theorem, which yields additional positive quantities on the boundary
hypersurfaces.
In particular,
the weights are sufficient to derive decay of the form $(\ref{fluxdecay})$, assuming
that the initial boundary integral is bounded.
Pointwise decay estimates of the form $(\ref{pointwise})$ can 
then be obtained by Sobolev inequalities, after commuting 
the equation with angular momentum operators
$\Omega$.

Turning to equation $(\ref{Lwe})$ on the Schwarzschild exterior, we have again
a timelike Killing vector field $\frac{\partial}{\partial t}$, and thus one 
immediately obtains conservation of the associated energy.
Applying as before the above method
to the vector field\footnote{defined with respect to suitably
normalized Eddington-Finkelstein advanced and retarded coordinates
$v$ and $u$} $K$ in the region $\{1\le t\le t_i\}$, 
we again obtain boundary hypersurface integrals with a sign,
and strong weights, in particular controlling
$t_i^{-2}$ times the energy density on $\{t= t_i\}\cap \{v\ge t_i\}\cap \{u\ge t_i\}$.
\[
\input{3perioxes.pstex_t}
\]
On the other hand, even after an additional integration by parts, 
the spacetime term arising from $\nabla^\alpha P_\alpha$ no longer 
vanishes. In the regions $r\le r_0$ and $r\ge R$, 
for certain constants $r_0$, $R$, however, 
this spacetime term has a good sign, and can be ignored.
On the other hand, 
in the region $r_0\le r\le R$, it turns out the spacetime integral arising from $K$ 
can be controlled by
$t$ times the spacetime integrals--summed--arising from vector fields $X_\ell$
of the form
\[
X_\ell=f_\ell\left(\frac{\pa}{\pa v}-\frac{\pa}{\pa u}\right),
\]
for a carefully chosen function $f_\ell$, applied to each spherical 
harmonic $\phi_\ell$, (see also \cite{bs} in connection with vector fields 
$X$). In fact, for this bound,
the vector fields $X_\ell$ must also be applied to angular derivatives
of $\phi$. This leads to loss of derivatives in the argument.\footnote{The necessity
of taking angular derivatives is related to the presence of the so-called
photosphere at $r=3M$.}
The boundary hypersurface integrals arising from $X_\ell$ 
are controlled in turn by
the total energy.
Putting the information together from these
two vector fields immediately yields
energy decay $(\ref{fluxdecay})$, but with power $-1$ in place of $-2$.

The boundary integrals arising from the sum of the $X_\ell$ identities, 
when suitably
localized to the future and past boundaries
of a dyadic characteristic rectangle $\{t_0\le t\le 1.1t_0\}\cap\{r_0\le r\le R\}$,
can in fact be bounded by $t_0^{-2}$ times the boundary
integrals arising from $K$ on constant $t=t_0$ and $t=1.1t_0$ hypersurfaces. 
Using this fact, we can iterate the procedure,
to obtain that the boundary integrals arising from $K$ are in fact bounded, and
thus that $(\ref{fluxdecay})$ holds as stated.

The above methods do not give good control
near the horizon $\mathcal{H}^+$.\footnote{This is not surprising,
as $K$ is normalized to $\mathcal{I}^+$.}
For this we need another estimate, which has
no analogue in Minkowski space. 
This estimate arises from applying a vector field of the form 
\[
Y\sim\frac{1}{1-\mu}\frac \partial{\partial u}
\]
in the characteristic rectangle depicted. The $v$ length of this rectangle
is chosen to be of the order of $t_i$. Note that $Y$ extends
regularly to the horizon $\mathcal{H}^+$. The boundary terms
arising from $Y$ are related to the energy that would be observed
by a local observer crossing the event horizon.\footnote{More precisely,
exactly the part of the energy not seen by $\frac{\partial}{\partial t}$.} 
The associated
spacetime integral
contains a term with a good sign, and terms that can be controlled by
the sum of the spacetime integrals arising from the $X_\ell$.
As a first step, one can show using the energy identities for
$\frac\partial{\partial t}$ and $X_\ell$, and the nature of the spacetime integral
arising from $Y$, that the boundary terms arising from $Y$ are uniformly bounded.
 Then one can go back and, using a pigeonhole
argument, extract from the term 
in the spacetime integral with a good sign
a constant-$v$ slice of the
rectangle such that the integral of the energy density as measured by a local
observer
is bounded by $t_i^{-1}$. Finally, one applies again
the energy identity of $Y$ in a subrectangle to
 obtain that the integral of the energy density measured by a local observer
is bounded by $t_i^{-1}$
on the segment $v=t_i$ depicted. One repeats the procedure to obtain
decay of $t_i^{-2}$.

Note that the procedure above for extracting decay 
for energy as measured by a local observer by means of a pigeonhole
argument applied to the spacetime integral arising from the energy 
identity for
$Y$ is the analytic manifestation in our technique
of the redshift effect described previously.
Here, the estimate is to be compared with the geometric
optics argument of the first diagram, but where the observers
$A$ and $B$ both cross the event horizon, with $B$ at advanced time
$t_i^{-1}$ later than $A$.

As in Minkowski space,
one obtains pointwise estimates $(\ref{pointwise})$ from the $L^2$ estimates
via the Sobolev inequality, after
commuting the equation with angular momentum operators. One should note however, that 
the necessary $L^2$ estimates near the horizon for the angular derivatives arise
from the decay rate of $t_i^{-2}$ for the energy flux measured by
$Y$. The 
boundary terms of $K$ and $\frac{\partial}{\partial t}$
do not contain angular derivatives on $\mathcal{H}^+\cup\mathcal{H}^-$.

Our use of the vector field  $Y$ is inspired by estimates in~\cite{dr1}.
In that paper, one could continue the iteration further to obtain
better decay rates than $(\ref{fluxdecay})$, $(\ref{pointwise})$. 
Here on the other hand, the weights on the boundary term
arising from the Morawetz identity give an upper bound to the amount of decay that
can be extracted. It would be interesting to explore  whether one can surpass
this barrier using additional techniques.

A final interesting aspect of our argument is that it does not require
inverting the Laplacian on initial data or appealing to the discrete isometries
of Schwarzschild. (In particular, the results here yield
an independent proof of the classical uniform boundedness
theorem~\cite{kw:lss} of Kay and Wald.)
It may be useful for non-linear applications to avoid techniques so heavily
dependent on the exact staticity. In this sense, the 
argument given here is perhaps more robust.

\section{Schwarzschild}
\label{Schwarzschild}
Let $(\mathcal{M},g)$ denote the maximally extended 
Schwarzschild spacetime with parameter $M>0$. This manifold is spherically
symmetric, i.e.~the group $SO(3)$ acts by isometry.
We recally briefly
the usual description of the global causal structure of $\mathcal{M}$,
via its so-called Penrose diagram. 
(We refer the reader to standard references, for instance~\cite{he:lssst}.)

For spherically symmetric spacetimes, recall that
Penrose diagrams are just the image of 
global bounded null coordinate systems on the Lorentzian quotient
$\mathcal{Q}=\mathcal{M}/SO(3)$ viewed as maps in the obvious way
from $\mathcal{Q}\to{\mathbb R}^{1+1}$. In the case of
Schwarzschild, the Penrose diagram is depicted below:
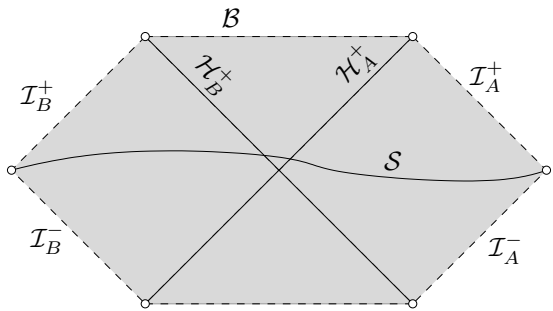
\[
\input{schwarzschild.pstex_t}
\]
The curve $\mathcal{S}$ depicts the projection
to $\mathcal{Q}$ of a particular choice
of complete Cauchy surface $\Sigma\subset \mathcal{M}$.
We will call the sets $J^-(\mathcal{I}^+_A)\cap J^+(\mathcal{I}^-_A)$
and $J^-(\mathcal{I}^+_B)\cap J^+(\mathcal{I}^-_B)$ the
exterior regions.\footnote{For an explanation of
the notation $J^-(\mathcal{I}^+)$, etc., 
and more about Penrose diagrams, see the appendix
 of~\cite{dr1}.} We call $\mathcal{H}^+_A$
the \emph{future event horizon} corresponding to the end $\mathcal{I}^+_A$
and $\mathcal{H}^+_B$ the future event horizon corresponding to $\mathcal{I}^+_B$.
We could also consider $\mathcal{H}^+_B$ as the past event horizon corresponding
to $\mathcal{I}^-_A$, and thus may denote it alternatively
by $\mathcal{H}^-_A$.

Defining a function
$r:\mathcal{Q}\to{\mathbb R}$
\begin{equation}
\label{arearadius}
r(q)=\sqrt{Area(q)/4\pi},
\end{equation}
the two exteriors are each covered by a coordinate system $(r,t)$
so that the metric $g$ may be written:
\begin{equation}
\label{Sch}
-\left(1-\frac{2M}r\right)dt^2+\left(1-\frac{2M}r\right)^{-1}
dr^2+r^2 d\sigma_{{\mathbb S}^2},
\end{equation}
where $d\sigma_{{\Bbb S}^2}$ denotes the standard metric on the unit sphere.
The vector field $\frac{\partial}{\partial t}$ is clearly timelike Killing.
These coordinates break down on $\mathcal{H}^+_A\cup\mathcal{H}^+_B$.
The Schwarzschild solution was originally understood as the spacetime
described by the expression $(\ref{Sch})$. As explained in the introduction,
the realization that this was actually just one of
the exterior regions of a larger spacetime, took
a surprisingly long time.\footnote{Note that the expression $(\ref{Sch})$
also describes the metric in the black
hole interior $J^+(\mathcal{H}^+_A)\cap J^-(\mathcal{H}^+_B)\setminus(
\mathcal{H}^+_B\cup\mathcal{H}^+_A)$. Here $\frac{\partial}{\partial t}$ is
spacelike Killing.}

The sphere $\mathcal{H}^+_B\cap \mathcal{H}^+_A$ is known as the \emph{bifurcation
sphere} of the event horizon. If we chose $\mathcal{S}$ to contain
$\mathcal{H}^+_B\cap\mathcal{H}^+_A$, then 
\[
\mathcal{S}'\doteq\mathcal{S}\cap J^-(\mathcal{I}^+_A)
\cap J^+(\mathcal{I}^-_A)
\]
is a Cauchy surface for the exterior region
$J^-(\mathcal{I}^+_A)
\cap J^+(\mathcal{I}^-_A)$. In particular, solutions of the linear wave
equations on  $J^-(\mathcal{I}^+_A)\cap J^+(\mathcal{I}^-_A)$
are determined by their data on $\mathcal{S}'$. 

The above means that if one is only interested in the behaviour
of solutions to $(\ref{Lwe})$ in $J^-(\mathcal{I}^+_A)\cap J^+(\mathcal{I}^-_A)$,
one can study the problem in any coordinate system defined globally in this region,
in particular, so-called Schwarzschild coordinates $(r,t)$. For convenience,
we shall in fact use a null coordinate system $(u,v)$,
which ``sends'' $\mathcal{H}^+_A$ to $u=\infty$ and $\mathcal{H}^+_B$ to 
$v=-\infty$. These coordinates, so-called Eddington-Finkelstein retarded
and advanced coordinates, will be described in the next section.

The reader should not think, however, that imposing such a coordinate
system removes the geometry from
this problem. 
For one must not forget that the correct assumptions on $\phi$ are those
expressable geometrically, not those that happen to look natural in the chosen
coordinate system.\footnote{For instance, one should not restrict
to $\phi$ of compact support on $\mathcal{S}'$, for this 
would mean that $\phi$ necessarily vanishes 
at $\mathcal{H}^+_B\cap\mathcal{H}^+_A$,
an actual sphere in the spacetime. See also the remarks in the 
Introduction.}
When written in local coordinates, these assumptions would be difficult 
to motivate without knowing the origin of the problem. Moreover, the same
can be said for the form of many of the techniques used here. In particular,
the form of the vector field $Y$ of Section~\ref{Ysec} is best understood by passing
to a new regular coordinate system on $\mathcal{H}^+$. In the coordinate system
chosen here, $Y$ appears asymptotically singular.

With these warnings in place, we turn to a description of two related coordinate
systems covering $J^-(\mathcal{I}^+_A)\cap J^+(\mathcal{I}^-_A)$.

\section{Eddington-Finkelstein and Regge-Wheeler coordinates}
Let $(r,t)$ denote the Schwarzschild coordinates of $(\ref{Sch})$. 
Define first the so-called Regge-Wheeler tortoise coordinate $r^*$
by\footnote{Coordinates $(r^*,t)$ are together known as
Regge-Wheeler coordinates. We have centred $r^*$ so as for
$r^*=0$ to correspond to $r=3M$. This is the so-called
\emph{photosphere}.}  
\[
r^*= r+2M\log(r-2M)-3M-2M\log M,
\]
and define retarded and advanced Eddington-Finkelstein coordinates
$u$ and $v$, respectively, by
\[
t=v+u
\]
and
\[
r^*=v-u.
\]
These coordinates turn out to be null: Setting
$\mu=\frac {2M}r$, 
the metric has the form
\[
-4(1-\mu)du dv+r^2d\sigma_{\mathbb S}^2.
\]

We shall move freely between the two coordinate systems $(r^*,t)$ and 
$(u,v)$ in this paper.
Note that in either, $J^-(\mathcal{I}_A^+)\cap J^-(\mathcal{I}_A^-)$ is covered
by $(-\infty,\infty)\times (-\infty,\infty)$.
By appropriately rescaling $u$ and $v$ to have finite range, one
can construct coordinates which are in fact regular on $\mathcal{H}^+$
and $\mathcal{H}^-$. By a slight abuse of language, one can parametrise the
future and past event horizons in our present
$(u,v)$ coordinate systems as $\mathcal{H}^+=\{(\infty,v)\}_{v\in[-\infty,\infty)}$ and 
$\mathcal{H}^-=\{(u,-\infty)\}_{u\in(-\infty,\infty]}$.

Finally, we collect various formulas for future reference:
\[
\mu=\frac {2M}r,
\]
\[
g_{uv}=(g^{uv})^{-1}=-2(1-\mu),
\]
\[
\partial_vr=(1-\mu),\qquad \partial_ur=-(1-\mu)
\]
\[
dt=dv+du, dr^*=dv-du,
\]
\[
\frac{\partial}{\partial t}=\frac12\left(\frac{\partial}{\partial v}
+\frac{\partial}{\partial
u}\right),
\]
\[
\frac{\partial}{\partial r^*}=\frac12\left(\frac{\partial}{\partial v}-\frac{\partial}{\partial
u}\right),
\]
\[
dVol_{\mathcal{M}}= r^2(1-\mu)\, du\, dv\, d\sigma_{{\mathbb S}^2},
\]
\[
dVol_{t=const}=r^2\sqrt{1-\mu}\, dr^*\, d\sigma_{{\mathbb S}^2},
\]
\[
\Box\psi=\nabla^\alpha\nabla_\alpha\psi= -(1-\mu)^{-1}\left(
\partial_t^2\psi-r^{-2}\partial_{r^*}(r^2\partial_{r^*}\psi)\right)
+\nabb^A\nabb_A\psi.
\]
Here $\nabb$ denotes the induced covariant derivative on the group orbit
spheres.

\section{The class of solutions}
Let $\mathcal{S}'$ denote the surface $\{t=1\}$ say in the exterior,
and let $\mathcal{S}$ be a Cauchy surface for
$\mathcal{M}$ such that $\mathcal{S}\cap J^-(\mathcal{I}^+)\cap J^+(\mathcal{I}^-)=
\mathcal{S}'$. Let $N$ denote the future-directed
unit normal to $\mathcal{S}$. 

We proceed to describe the solutions $\phi:\mathcal{M}\to{\mathbb R}$ of 
the wave
equation
\begin{equation}
\label{waveeq}
\Box_g\phi=0
\end{equation}
on $\mathcal{M}$, which we shall consider in this paper. Given
$\phi$, $ N\phi$ on $\mathcal{S}$,
define the quantities
\begin{eqnarray}
\nonumber
\label{tf0}
\bar{E}_0&=&\sum_{i=0}^2
\int\limits_{-\infty}^{\infty}\int\limits_{{\mathbb S}^2}
r^2(1-\mu)^{-\frac12}\left((\partial_t (r^i\nabb^i\phi))^2+
(\partial_{r^*} (r^i\nabb^i\phi))^2\right.\\
&&\hbox{}\left.+
(1-\mu) |\nabb r^i\nabb^{i} \phi|^2\right)
(1,r^*,\sigma_{{\Bbb S}^2}) dr^* 
d\sigma_{{\mathbb S}^2},
\end{eqnarray}
\begin{eqnarray}
\nonumber
\label{tf}
\bar{E}_1&=&\sum_{i=0}^{2}\int
\limits_{-\infty}^\infty\int\limits_{{\mathbb S}^2}
r^2\left(u^2(\partial_u (r^i\nabb^{i}\phi))^2+v^2(\partial_v (r^i\nabb^{i}
\phi))^2\right.\\
\nonumber
&&\hbox{}\left.+(1-\mu)(u^2+v^2)|\nabb (r^i\nabb^{i}\phi)|^2\right) 
(1,r^*,\omega)
dr^*d\sigma_{{\mathbb S}^2}\\
\nonumber
&&\hbox{}+\sum_{i=0}^3
\int\limits_{-\infty}^{\infty}\int\limits_{{\mathbb S}^2}
r^2 \left ((\partial_t (r^i\nabb^i\phi))^2+
(\partial_{r^*} (r^i\nabb^i \phi))^2\right.\\
&&\hbox{}+\left.(1-\mu) 
|\nabb (r^i \nabb^i \phi)|^2\right)
(1,r^*,\omega) dr^*  d\sigma_{{\mathbb S}^2}
\end{eqnarray}
\begin{eqnarray}
\nonumber
\label{tf2}
\bar{E}_2&=&\sum_{i=0}^{4}
\int\limits_{-\infty}^\infty\int\limits_{{\mathbb S}^2}
r^2\left(u^2(\partial_u (r^i\nabb^{i}\phi))^2+v^2(\partial_v (r^i\nabb^{i}
\phi))^2\right.\\
\nonumber
&&\hbox{}\left.
+(1-\mu)(u^2+v^2)|\nabb (r^i\nabb^{i}\phi)|^2\right)
(1,r^*,\omega)dr^*d\sigma_{{\mathbb S}^2}\\
\nonumber
&&\hbox{}+\sum_{i=0}^5
\int\limits_{-\infty}^{\infty}\int\limits_{{\mathbb S}^2}
r^2 \left ((\partial_t (r^i\nabb^i\phi))^2+
(\partial_{r^*} (r^i\nabb^i \phi))^2\right.\\
\nonumber
&&\hbox{}\left.+(1-\mu) 
|\nabb (r^i \nabb^i\phi)|^2\right)
(1,r^*,\omega) dr^*  d\sigma_{{\mathbb S}^2}\\
\nonumber
&&\hbox{}+\sum_{i=0}^2
\int\limits_{-\infty}^{\infty}\int\limits_{{\mathbb S}^2}
r^2(1-\mu)^{-\frac12}\left((\partial_t (r^i\nabb^i\phi))^2+
(\partial_{r^*} (r^i\nabb^i\phi))^2\right.\\
&&\hbox{}\left.
+(1-\mu) |\nabb (r^i\nabb^i\phi)|^2\right)
(1,r^*,\omega)\cdot dr^* d\sigma_{{\mathbb S}^2}.
\end{eqnarray}
Our weakest result, namely, the uniform boundedness, requires 
the boundedness of $\bar{E}_0$. Our energy decay result will require
the boundedness of $\bar{E}_1$
and our full pointwise decay 
results will require the boundedness of $\bar{E}_2$. 

We note that the boundedness of $\bar{E}_0$ follows from the statement
that $\phi$ is $C^3$, $N\phi$ is $C^2$ on $\mathcal{S}$, 
and that $\phi$ and decays suitably
at spacelike infinity, for instance, if $\phi$ vanishes identically in a
neighborhood of $i^0$. Similarly, the boundedness of $\bar{E}_1$ follows
for suitably decaying $C^4$ $\phi$ and $C^3$ $N\phi$, and finally
the boundedness of $\bar{E}_2$ follows from $C^6$ $\phi$, etc.
There is no assumption of the vanishing of $\phi$ on $\mathcal{H}^+\cap
\mathcal{H}^-$. 

We will state a coordinate version of Theorem~\ref{main}
in Section~\ref{cvsec}. As this theorem will refer to the energy flux
defined by the Killing vectorfield $\frac{\partial}{\partial t}$, we will
first need some general results regarding conservation laws.
These will be given in the next two sections.

\section{Conservation laws}
As discussed in the introduction, the results of this paper will rely
on estimates of energy-type. Such estimates arise naturally in view of
the Lagrangian structure of the theory. We review briefly here.

In general coordinates,
the energy-momentum tensor for $\phi$  is given by
\[
T_{\alpha\beta}=\partial_\alpha\phi\partial_\beta\phi-\frac12g_{\alpha\beta}g^{\gamma\delta}
\partial_\gamma\phi\partial_\delta\phi.
\]
This is divergence free, i.e.~we have
\begin{equation}
\label{divfree}
\nabla^\alpha T_{\alpha\beta}=0.
\end{equation}

For the null coordinates we have defined, we compute the components
\[
T_{uu}=(\partial_u\phi)^2,
\]
\[
T_{vv}=(\partial_v\phi)^2,
\]
\[
T_{uv}=-\frac12g_{uv}|\nabb\phi|^2=(1-\mu)|\nabb\phi|^2.
\]
Let $V^\alpha$ denote an arbitrary vector field.
Let $n^\alpha$ denote the  normal to a hypersurface\footnote{Note the convention in
Lorentzian geometry}. Let $\pi_V^{\alpha\beta}$ denote the deformation tensor of $V$,
i.e.,
\begin{equation}
\label{defdef}
\pi_V^{\alpha\beta}=\frac12(\nabla^\alpha V^\beta+\nabla^\beta V^\alpha).
\end{equation}
We will denote in what follows the tensor $\pi^{\alpha\beta}_V$ 
just by $\pi^{\alpha\beta}$.
In local coordinates we have the following expression:
\begin{eqnarray*}
T_{\alpha\beta}\pi ^{\alpha\beta}&=&\frac{1}{4(1-\mu)}\left(
(\partial_u\phi)^2\partial_v(V_v(1-\mu)^{-1})
				+(\partial_v\phi)^2\partial_u(V_u(1-\mu)^{-1})\right.\\
				&&\hbox{}\left.+|\nabb\phi|^2(\partial_u 
V_v+\partial_v V_u)
					\right)
			-\frac1{2r}(V_u-V_v)(|\nabb\phi|^2-\phi^\alpha\phi_\alpha).
\end{eqnarray*}

Let $\mathcal{S}$ be a region bounded to the future and past by two hypersurfaces
$\Sigma_1$ and $\Sigma_0$, respectively. Let 
$P^\alpha=g^{\alpha\beta}T_{\beta\delta}X^\delta$.
The divergence theorem together with $(\ref{divfree})$ and $(\ref{defdef})$ gives
\begin{equation}
\label{divthe}
\int_{\mathcal{S}} T_{\alpha\beta}\pi^{\alpha\beta} dVol_{\mathcal{S}}=
\int_{\Sigma_1} g_{\alpha\beta}P^\alpha
n^\beta dVol_{\Sigma_1}-
\int_{\Sigma_0} g_{\alpha\beta}P^\alpha n^\beta dVol_{\Sigma_0}.
\end{equation}

\section{Conservation of energy}
Let us apply the above to the vector field $\frac{\partial}{\partial t}$. As this
is Killing, by $(\ref{defdef})$ and $(\ref{divfree})$, the 
associated vector field $P$ is divergence free, so there is no
space-time term.
The (negative of the) boundary terms on constant
$t$-hypersurfaces are:
\begin{eqnarray*}
E_{\phi}(t_i)&\doteq&
\int\limits_{-\infty}^\infty\int\limits_{{\mathbb S}^2}\frac1{\sqrt{1-\mu}}
\left(\frac14T_{vv}+\frac14T_{uu}+\frac12T_{uv}\right)(r^*,t_i,\omega)\cdot
r^2\sqrt{1-\mu}\,
dr^* d\sigma_{{\mathbb S}^2}\\
&=&
\int\limits_{-\infty}^\infty\int\limits_{{\mathbb S}^2}
\frac1{4\sqrt{1-\mu}}((\partial_v\phi)^2+(\partial_u\phi)^2+2(1-\mu)|\nabb\phi|^2) 
\cdot r^2\sqrt{1-\mu}\,dr^* 
d\sigma_{{\mathbb S}^2}\\
&=&
\int\limits_{-\infty}^{\infty}\int\limits_{{\mathbb S}^2}
\frac1{2\sqrt{1-\mu}}((\partial_t\phi)^2+(\partial_{r^*}\phi)^2+(1-\mu)|\nabb\phi|^2) 
\cdot r^2\sqrt{1-\mu}\,dr^* 
d\sigma_{{\mathbb S}^2}
\end{eqnarray*}
on constant-$u$ hypersurfaces are given by:
\begin{eqnarray*}
F_{\phi}(\{u\}\times[v_1,v_2])&\doteq&
\int_{v_1}^{v_2}\int_{{\mathbb S}^2}\frac1{4(1-\mu)}(T_{vv}+T_{uv})\cdot r^2(1-\mu)
dv d\sigma_{{\mathbb S}^2}\\
&=&\int_{v_1}^{v_2}\int_{{\mathbb S}^2}
\frac1{4(1-\mu)}((\partial_v\phi)^2+(1-\mu)|\nabb\phi|^2)\cdot r^2(1-\mu)
dv d\sigma_{{\mathbb S}^2}
\end{eqnarray*}
and on constant-$v$ hypersurfaces are given by:
\begin{eqnarray*}
F_{\phi}([u_1,u_2]\times\{v\})&\doteq&
\int_{u_1}^{u_2}\int_{{\mathbb S}^2}\frac1{4(1-\mu)}(T_{uu}+T_{uv})\cdot r^2(1-\mu)
du d\sigma_{{\mathbb S}^2}\\
&=&\int_{u_1}^{u_2}\int_{{\mathbb S}^2}
\frac1{4(1-\mu)} ((\partial_u\phi)^2+(1-\mu)|\nabb\phi|^2)\cdot r^2(1-\mu)
du d\sigma_{{\mathbb S}^2}.
\end{eqnarray*}

The identity $(\ref{divthe})$ applied to the above shows that 
\[
E_\phi(t_0)=E_\phi(1)\doteq E_\phi
\]
and that the fluxes satisfy
\[
F_{\phi}([u_1,u_2]\times\{v\})\le E_\phi,
\]
\[
F_{\phi}(\{u\}\times[v_1,v_2])\le E_\phi.
\]
In particular, we can define a function
$\varpi_\phi$ on $\mathcal{Q}\cap \overline{J^-(\mathcal{I}^+)\cap 
J^+(\mathcal{I}^-)}$ 
by
\begin{equation}
\label{pidef}
\varpi_\phi(u,v)\doteq F_{\phi}([u,\infty]\times\{v\})
+F_{\phi}(\{\infty\}\times[v,\infty]).
\end{equation}
It is clear that the bound
\begin{equation}
\label{globalenergybound}
\varpi_\phi \le 2E_{\phi},
\end{equation}
follows immediately.

\section{Coordinate version of the main theorem}
\label{cvsec}
\begin{theorem}
\label{cvmt}
Let $\phi_0(r^*,\omega)$, $\phi_1(r^*,\omega)$ be functions such that the quantity
$\bar{E}_0$ of $(\ref{tf0})$ is bounded\footnote{in the obvious sense,
i.e.~when $\phi$ is replaced by $\phi_0$ and $\partial_t\phi$ is replaced
by $\phi_1$ when evaluating $(\ref{tf0})$}, 
and let
$\phi$ be the unique solution of $(\ref{Lwe})$
on $J^-(\mathcal{I}^+)\cap J^+(\mathcal{I}^-)$ 
with $\phi(r^*,1,\omega)=\phi_0(r^*,\omega)$, 
$\partial_t\phi(r^*,1,\omega)=\phi_1(r^*,\omega)$.
Let $\varpi_\phi$ be as defined in $(\ref{pidef})$. 
Then there exists a universal constant $C$ 
such that
\begin{equation}
\label{Cub}
|\phi|\le C\bar{E}_0 r^{-\frac12}.
\end{equation}
If the quantity  $\bar{E}_1$ of $(\ref{tf})$ is bounded, then
\begin{equation}
\label{Cfb}
\varpi_\phi\le C\bar{E}_1(v_+^{-2}+u_+^{-2}).
\end{equation}
If the quantity $\bar{E}_2$ of $(\ref{tf2})$ is bounded, then 
\begin{equation}
\label{Cpb1}
|\phi|\le C\bar{E}_2v_+^{-1}
\end{equation}
for all $(u,v,\omega)$, 
while
\begin{equation}\label{Cpb2}
|r\phi|\le C_{\hat{R}}\bar{E}_2 (1+|u|)^{-\frac12}
\end{equation}
for all $r\ge \hat{R}>2M$, $t\ge 1$.
\end{theorem}

In view of our previous remarks, it is clear that Theorem~\ref{cvmt}
 implies Theorem~\ref{main}.

\begin{remark}
The number of derivatives required in the definitions of $\bar{E}_1, \bar{E}_2$
can be reduced by a slight refinement of the analysis in Section \ref{X}. 
We shall not pursue this here. 
\end{remark}

\section{The vector fields $X_{\ell}$}\label{X}
In this section we shall 
define, for each spherical harmonic $\phi_{\ell}$, a 
vector field $X_{\ell}$ by
\begin{equation}
\label{thisform}
X_{\ell}=-\frac12 
f_{\ell}\frac{\partial}{\partial u}+\frac12f_{\ell}\frac{\partial}{\partial v}
=f_{\ell}\frac{\partial}{\partial r^*},
\end{equation}
for some function $f_{\ell}=f_{\ell}(r^*)$ to be determined later.
We shall show that
the function $f_{\ell}$ can be chosen so as for the spacetime term corresponding
to $X_{\ell}$ to be positive, and so as for the boundary terms arising to be controlled
by the usual (conserved)
$\frac{\partial}{\pa t}$-energy, 
after application of a Hardy inequality. 
(See Proposition~\ref{withhardy}
from the next section.)
These $X_{\ell}$-energy identites can then be summed so as to yield an identity
relating a positive spacetime integral and boundary terms controlled
by the usual $\frac{\pa}{\pa t}$-energy. (See Section~\ref{postermsec}.)

\subsection{Identities}
We first collect various identities for vector fields of the form $(\ref{thisform})$
for a function $f$.
For functions of $r^*$, let $'$ here denote $\frac{d}{d r^*}$.
The spacetime integral given 
by the left hand side of 
$(\ref{divthe})$ in a region $\mathcal{R}=\{t_0\le t\le t_1\}$ 
is:
\begin{eqnarray}
\label{nova2}
\nonumber
\hat{I}^X_{\phi}(\mathcal{R})&=&\int_{\mathcal{R}}T_{\alpha\beta}
\pi^{\alpha\beta} 
dVol\\
\nonumber
&=&\int_{t_0}^{t_1}\int_{-\infty}^{\infty}\int_{{\mathbb S}^2}
\left(\frac{f'(\partial_{r^*}\phi)^2}{1-\mu}
+\frac12|\nabb\phi|^2\left(\frac{2-3\mu}r\right)f\right.\\
&&\hbox{}\left.-\frac14\left(2f'+4\frac{1-\mu}r f\right)\phi^\alpha
\phi_\alpha\right)\cdot r^2(1-\mu)dt\,dr^*\,d\sigma_{{\mathbb S}^2}
\end{eqnarray}
while the boundary terms are given on a constant $t$-hypersurfaces by:
\[
\hat{E}^X_{\phi}(t_i)=
\int\limits_{-\infty}^\infty\int\limits_{{\Bbb S}^2} 
f\pa_t \phi \, \pa_{r_*} \phi(t_i,r^*,\omega)
\,  r^2\sqrt{1-\mu}\, dr_*\, d\sigma_{{\mathbb S}^2}.
\]
We have the identity
\begin{equation}
\label{origXid}
\hat{I}^X_{\phi}(\mathcal{R})=\hat{E}^X_{\phi}(t_1)-\hat{E}^X_{\phi}(t_0).
\end{equation}

To produce an identity with terms for which we can control the signs,
we wish to replace the spacetime integral $\hat{I}^X_{\phi}$ with a new
integral obtained by applying Green's theorem to its last term.
Defining
\begin{eqnarray}
\label{XLdef}
\nonumber
I^X_{\phi}(\mathcal{R})&=&
\int_{t_0}^{t_1}\int_{-\infty}^{\infty}\int_{{\mathbb S}^2}\left(
		\frac{f'}{1-\mu}(\partial_{r^*}\phi)^2
		+|\nabb\phi|^2\left(\frac{\mu'}{2(1-\mu)}+
		\frac{1-\mu}{r}\right)f\right.\\
		&&\left.\hbox{}-\frac 14 \left (\Box \left(f'+2\frac{1-\mu}rf\right) 
		\right) \phi^2
			\right)\,r^2\,(1-\mu)\,dt\, dr^*\, d\sigma_{{\mathbb S}^2},
\end{eqnarray}
we have by Green's theorem that 
\begin{eqnarray*}
I^X_{\phi}(\mathcal{R})&=&  \hat{I}^X_{\phi}(\mathcal{R})
+ \int\limits_{-\infty}^\infty\int\limits_{{\Bbb S}^2} 
\frac12\left(f'+2\frac{1-\mu}r\right)\pa_t\phi\, \phi(t_1,r^*,\omega)
\,  r^2\, dr^*\, d\sigma_{{\mathbb S}^2} \\
&&\hbox{}-\int\limits_{-\infty}^\infty\int\limits_{{\Bbb S}^2} 
\frac12\left(f'+2\frac{1-\mu}r\right)\pa_t\phi\, \phi(t_0,r^*,\omega)
\,  r^2\, dr^*\, d\sigma_{{\mathbb S}^2},
\end{eqnarray*}
and thus we have the identity
\[
I^X_{\phi}(\mathcal{R})=E^X_{\phi}(t_1)-E^X_{\phi}(t_0)
\]
where 
\begin{equation}
\label{Xbtdef}
E^X_{\phi}(t_i)=\hat{E}^X_{\phi}(t_i)+ \int\limits_{-\infty}^\infty\int\limits_{{\Bbb S}^2} 
\frac12\left(f'+2\frac{1-\mu}r\right)\pa_t\phi\, \phi(t_i,r^*,\omega)
\,  r^2\sqrt{1-\mu}\, dr^*\, d\sigma_{{\mathbb S}^2}.
\end{equation}

Finally, let us compute the expression with the d'Alambertian on the
right hand side of $(\ref{XLdef})$.
Since, for a function $\psi$ dependent only on $r$, we have
$\Box\psi = (1-\mu)^{-1}\psi '' +\frac2r\psi'$, 
it follows that 
\begin{eqnarray*}
\Box\left(f'+2\frac{1-\mu}rf\right)&=&
\frac{1}{1-\mu}f'''+\frac2rf''+\frac2rf'' \\
&&\hbox{}+\frac{2}{1-\mu}\left(\left(\frac{1-\mu}{r}\right)''f-
2f'\frac{(1-\mu)^2}{r^2}
-2\frac{f'}r\mu'\right) \\
&&\hbox{}+\frac{4}{r^2}(1-\mu)f'-\frac{4}{r^3}(1-\mu)^2f-
\frac{4}{r^2}\mu'f.
\end{eqnarray*}
In view of the relations
\[
\left(\frac{1-\mu}{r}\right)'=-\frac{(1-\mu)^2}{r^2}-\frac{\mu'}r
\]
and
\[
\left(\frac{1-\mu}{r}\right)''=2\frac{(1-\mu)^3}{r^3}+
						3\frac{(1-\mu)\mu'}{r^2}
						-\frac{\mu''}r,
\]
we obtain the expression
\begin{eqnarray}
\label{theexpression}
\nonumber
\Box\left(f'+2\frac{1-\mu}rf\right)&=& \frac{1}{1-\mu}f'''+\frac 4rf''-\frac{4\mu'}{r(1-\mu)}f'
	\\
	&&\hbox{}+\frac{2}{(1-\mu)r}\left(\frac{\mu'(1-\mu)}r-\mu''\right)f.
\end{eqnarray}

\subsection{The choice of $f_{\ell}$ for $\ell\ge1$}
In this section, we shall choose $f_{\ell}$ for the higher spherical harmonics
$\ell\ge 1$.

Our goal is to make the spacetime 
integral $I^{X_{\ell}}_{\phi_\ell}(\mathcal{R})$ positive,
and the boundary terms $E^{X_{\ell}}_{\phi_{\ell}}$ 
controllable by 
the energy $E_{\phi_{\ell}}$. 
Thus, one would like the function
$f_{\ell}$ to be bounded, and, in view of the first term in 
$(\ref{XLdef})$,
 $f'_{\ell}$ should be positive. For this to be the case,
however, $f'''_{\ell}$ 
must become positive in a neighborhood of the horizon.
Examining, however, $(\ref{XLdef})$ and $(\ref{theexpression})$,
one sees that this term enters 
in $I^{X_\ell}_{\phi_\ell}$ 
with the wrong sign, and, what is more,
it is multiplied by $(1-\mu)^{-1}$, and thus seems to dominate the
only term with the correct sign to cancel it, namely $f''_\ell$.

To generate a term which can indeed cancel this ``bad term'', 
we must borrow from
the term in $(\ref{XLdef})$
with $\partial_{r^*}\phi$. This is possible as follows:
For any $C^1$ function $\beta=\beta(r^*)$ decaying to $0$ as $r^*\to\pm\infty$, 
we note 
the following identity:
\begin{align*}
&\int_{-\infty}^\infty \frac{f'}{1-\mu}(\partial_{r^*}\phi)^2r^2 (1-\mu) dr^*
=
\int_{-\infty}^\infty \frac{f'}{1-\mu}(\partial_{r^*}\phi+\beta\phi)^2r^2 (1-\mu) dr^*\\
&\qquad+
\int_{-\infty}^\infty\phi^2\left(-\frac{f'}{1-\mu}\beta^2
+\frac{f''}{1-\mu}\beta +\frac{f'}{1-\mu}\beta'+
	\frac2rf'\beta
	\right)r^2(1-\mu) dr^*.
\end{align*}
Let us set $M=1$ (i.e.~$\mu=\frac2r$) and, for a 
sufficiently large constant 
 $\alpha$ to be determined below, let $x$ denote the coordinate 
\[
x=r^*-\alpha-1.
\]
Define
\[
\beta=\frac{1-\mu}{r}-\frac {x}{\alpha^2+x^2}, \qquad 
\beta'=-\frac{(1-\mu)^2}{r^2} -\frac{\mu'}r -\frac {1}{1+x^2}+
\frac {2x^2}{(\alpha^2+x^2)^2}
\]
so that
\begin{eqnarray*}
\beta'-\beta^2+\frac{2}{r}(1-\mu)\beta=
-\frac{\mu'}r-\frac{\alpha^2}{(\alpha^2+x^2)^2}.
\end{eqnarray*}
We may thus write
\begin{align*}
I_{\phi_\ell}^{X_\ell}(\mathcal{R})
=&\int\limits_{t_0}^{t_1}
\int\limits_{-\infty}^{\infty}\int\limits_{{\mathbb S}^2}\left(
\frac{f_{\ell}'}{1-\mu}\left(\partial_{r^*}\phi_\ell
+\left(\frac {1-\mu}r-\frac {x}{\alpha^2+x^2}\right)
\phi_\ell\right)^2+
\frac{2-3\mu}{2r} \, f_\ell\, |\nabb\phi_\ell|^2\right.\\ 
&-\frac14\frac1{1-\mu}\left({f'''_\ell}+ 
\frac{4 f''_\ell x}{\alpha^2+x^2}+\frac{4\alpha^2 
f'_\ell}{(\alpha^2+x^2)^2}\right)\phi_\ell^2\\
&\left.-
\frac{\mu f_\ell}{2r^3}\left(4\mu -3\right) \phi_\ell^2\right)
\, r^2\, (1-\mu)\, dt\, dr^*\, d\sigma_{{\mathbb S}^2}.
\end{align*}

For each spherical harmonic number $\ell$, in view 
of the relation
\[
 \int_{{\mathbb S}^2}|\nabb\phi_{\ell}|^2r^2d\sigma_{{\mathbb S}^2}
 =\int_{{\mathbb S}^2}\frac{(\ell)(\ell+1)}{r^2}
   (\phi_{\ell})^2r^2d\sigma_{{\mathbb S}^2},
\]
we may rewrite
\begin{align}
\nonumber
\label{kovta}
I_{\phi_{\ell}}^{X_{\ell}}(\mathcal{R})
=&\int\limits_{t_0}^{t_1}\int\limits_{-\infty}^{\infty}\int\limits_{{\mathbb S}^2}
\left(\frac{f_\ell'}{1-\mu}\left (\partial_{r^*}\phi_\ell+
\left(\frac {1-\mu}r-\frac {x}{\alpha^2+x^2}\right)
\phi_\ell\right)^2\right.\\
\nonumber
&\left.-\frac14\frac 1{1-\mu}\left({f_\ell'''}+ 
\frac {4 f_\ell'' x}{\alpha^2+x^2}+
\frac {4\alpha^2 f_\ell'}{(\alpha^2+x^2)^2}\right)
\phi_\ell^2
\right.\\
&\left.+\left (\ell(\ell+1) \frac {2-3\mu}{2r^3}+ \frac {3-4\mu}{r^4}\right)f_\ell
\phi_\ell^2
\right)\, r^2\, (1-\mu)\, dt\, dr^*\, d\sigma_{{\mathbb S}^2}.
\end{align}
The expression multiplying $f_\ell\phi_\ell^2$ in the last term above
vanishes at a unique value of $r^*$. Denote this by
$-\gamma_\ell$. Note that
\[
-4/3\le  r^*(8/3)\le -\gamma_\ell \le r^*(3)=0.
\]
We may now define $f_\ell=f_\ell(r^*)$ by setting
\[
f_{\ell}(\gamma(\ell))=0,
\] 
\[
(f_{\ell})'=\big (\alpha^2+x^2\big )^{-1}.
\]

Let us drop the $\ell$ in what follows.
We have
\[
f''=-\frac{2 x }{\big (\alpha^2+x^2\big )^{2}},
\]
\[
f'''=\frac{8x^2}{\big (\alpha^2+x^2\big )^{3}}-
\frac{2}
{\big (\alpha^2+x^2\big )^{2}}.
\]
Denoting by
\[
F= -\frac1{4(1-\mu)}\left({f'''}+ 
 \frac {4 f'' x}{\alpha^2+x^2}+\frac {4f'\alpha^2}{(\alpha^2+x^2)^2}
\right),
\] 
we compute
\[
F=\frac{1}{2(1-\mu)}
\, \frac{x^2 - \alpha^2} 
{\big (\alpha^2+x^2\big )^{3}}.
\]
Note that $F$ is nonnegative for $x\le-\alpha$, and $x\ge\alpha$.

We would like to show that for a sufficiently large $\alpha$, independent of $\ell$,
we can dominate the second term in
$(\ref{kovta})$ pointwise 
by the last term, i.e.~we want to show
\begin{equation}
\label{autopou9elw}
-F\le \left (\ell(\ell+1) \frac {2-3\mu}{2r^3}+ \frac {3-4\mu}{r^4}\right)
 f.
\end{equation}
In view of the sign of $F$, it suffices to consider $-\alpha<x<\alpha$.

For $-\alpha<x<\alpha$ then, we estimate
$$
f=\int_{-\alpha-1- \gamma}^x f'\ge \frac {x+\alpha+1}{2\alpha^2+2\alpha+1}.
$$
To show $(\ref{autopou9elw})$, it suffices then to show
\begin{equation}
\label{newin}
\left (\ell(\ell+1) \frac {2-3\mu}{2r^3}+ \frac {3-4\mu}{r^4}\right)
 \frac {x+\alpha+1}{2\alpha^2 +2\alpha+1}\ge
\frac{1}{2(1-\mu)}
\, \frac{ \alpha^2-x^2} 
{\big (\alpha^2+x^2\big )^{3}}\, .
\end{equation}

Consider first the region $-\alpha< x\le -2\alpha/3$. 
For $\alpha$ sufficiently large, we have
$$
\frac{1}{2(1-\mu)}
\, \frac{ \alpha^2-x^2} 
{\big (\alpha^2+x^2\big )^{3}}\le \frac{3\sqrt 2}{2}
\, \frac{ \alpha+x} 
{\big (\frac {13}9\alpha^2\big )^{5/2}}\, .
$$
On the other hand, since for $x\le -2\alpha/3$ we have, say  $r\le 5\alpha/12$, 
and 
since
$$
r-3=\int_{0}^{r^*} (1-\mu)\ge \frac 13 r^*,
$$
we have $(2-3\mu)(r^*)\ge 1/5$ for $r^*\ge 1$. Then for $\ell\ge 1$,
$$
\ell(\ell+1) \frac {2-3\mu}{2r^3}
 \frac {x+\alpha+1}{2\alpha^2 +2\alpha+1}\ge  
 \frac {x+\alpha}{\alpha^5},
$$
for sufficiently large $\alpha$.
This gives $(\ref{newin})$ for the region considered.

Consider now the region $-2\alpha/3\le x< \alpha$. Note that 
as $\alpha\to\infty$, we have  $r\sim x+\alpha$, $\mu\sim 0$ in this region.
Thus, we have
$$
\ell(\ell+1) \frac {2-3\mu}{2r^3}
 \frac {x+\alpha+1+\gamma}{2\alpha^2 +2\alpha+1}\sim   
 \frac{\ell(\ell+1)}2 \frac {x+\alpha}{(x+\alpha)^3 \alpha^2},
$$
while 
$$
\frac{1}{2(1-\mu)}
\, \frac{ \alpha^2-x^2} 
{\big (\alpha^2+x^2\big )^{3}}\sim \frac 12 \frac{ (\alpha-x)(\alpha+ x)} 
{\big (\alpha^2+x^2\big )^{3}}.
$$
To show $(\ref{newin})$ for sufficiently large choice of $\alpha$,
it suffices then to show the bound
\begin{equation}
\label{<1}
\frac {(\alpha-x)(x+\alpha)^3\alpha^2}{2(\alpha^2+x^2)^3}<1.
\end{equation}

For $x<0$, $(\ref{<1})$ is immediate from 
$$
\frac {(\alpha-x)\alpha^5}{2(\alpha^2+x^2)^3}\le 
\frac {\sqrt {2(\alpha^2+x^2)}\alpha^5}{2(\alpha^2+x^2)^3}\le 2^{-\frac 12}.
$$
For $x\ge 0$, we have on the one hand
$$
\frac {(\alpha-x)(x+\alpha)^3\alpha^2}{2(\alpha^2+x^2)^3}\le 
\frac {(x+\alpha)^3\alpha^3}{2(\alpha^2+x^2)^3}\le \frac 12\left (\frac {\alpha x+\alpha^2}
{x^2+\alpha^2}\right)^3.
$$
On the other hand,
$$
\alpha x+\alpha^2 \le \frac q2 x^2 +\left(1+\frac 1{2q}\right) \alpha^2.
$$
Set 
$$
q=2+\frac 1q, \qquad q^2-2q-1=0.
$$
Then $q=1+\sqrt 2$ and 
$$
\alpha x+\alpha^2\le \frac 12(1+\sqrt 2) (\alpha^2+x^2).
$$
The bound $(\ref{<1})$ then follows from the inequality
$$
\big((1/2)(1+\sqrt 2)\big)^3< 2.
$$

\subsection{The case $\ell=0$}
For $\ell=0$, we have the identity
$$
2T_{\alpha\beta}\pi^{\alpha\beta}= \frac{f'_0}{(1-\mu)} \left 
({(\partial_t\phi_0)^2}
					+{(\partial_{r^*}\phi_0)^2}\right) 
					+ \frac {2}r \left 
					((\pa_t\phi_0)^2-(\pa_{r^*}
					\phi_0)^2\right).
$$
Applying $(\ref{origXid})$ with the sharp cut-off function
 $f_0=\chi_{(-\infty,R^*)}$, so that 
$f_0'=-\delta(r^*-R^*)$, we obtain
\begin{align*}
&\int\limits_{t_0}^{t_1}  
\int\limits_{{\Bbb S}^2} \frac{1}{(1-\mu)} \left ({(\partial_t\phi_0)^2} +
(\partial_{r^*}\phi_0)^2\right ) (t,R^*,\omega) r^2 (1-\mu) 
d\sigma_{{\mathbb S}^2} dt\\ 
&\qquad+\int\limits_{t_0}^{t_1} \int\limits_{-\infty}^{R^*}
 \int\limits_{{\Bbb S}^2} \frac{2}{r} 
\left ( (\pa_{r^*}\phi_0)^2- (\pa_t\phi_0)^2\right) 
 r^2 (1-\mu) d\sigma_{{\mathbb S}^2} dr^* dt
 \le 4 E_{\phi_0}.
\end{align*}
Let us denote 
$$
F(r^*)= \int\limits_{t_0}^{t_1} \int\limits_{-\infty}^{r^*}
 \int\limits_{{\Bbb S}^2} \frac{2}{r} (\pa_t\phi_0)^2 r^2 (1-\mu) 
d\sigma_{{\mathbb S}^2} 
d\rho dt.
$$
Then, in particular,
$$
\frac r2 \frac 1{1-\mu} F'(r^*) \le F(r^*) + 4 E_{\phi_0}.
$$
Hence,
$$
\left (e^{-\int_{-\infty}^{r^*}2r^{-1} (1-\mu)} F(r^*)\right)' \le 4 
e^{-\int_{-\infty}^{r^*} 2r^{-1}(1-\mu)} 2r^{-1}(1-\mu) E_{\phi_0}.
$$
Note that $\int_{-\infty}^{r^*}2r^{-1}(1-\mu)dr^*=\log\mu^{-2}$.
 Also observe that $F(-\infty)=0$. Therefore, integrating 
we obtain 
$$
\mu^{2}
F(r^*)\le 4 E_{\phi_0} \int_{2M}^r \mu^{2} 
d\tilde{r}\le 4 E_{\phi_0},
$$
and thus,
$$
F(r^*) \le 4 \mu^{-2} E_{\phi_0}.
$$
It now follows that 
$$
\int\limits_{t_0}^{t_1}  
\int\limits_{{\Bbb S}^2} \frac{1}{(1-\mu)} \left ({(\partial_t\phi)^2} +
(\partial_{r^*}\phi)^2\right ) (t,r^*,\omega) 
r^2 (1-\mu) d\sigma dt \le 4 
\mu^{-2} E_{\phi_0}.
$$
Multiplying by $\mu^{2}$ and 
averaging over $r^*$ we obtain 
\begin{eqnarray}
\nonumber
\label{thedefi}
\hskip -3pc\tilde{I}^{X_0}_{\phi_0}&\doteq&\int\limits_{t_0}^{t_1} \int\limits_{-\infty}^\infty
 \int\limits_{{\Bbb S}^2} \frac{4\mu^{2}}{(1-\mu)(1+|r^*|)^{1+}} 
 \left({(\partial_t\phi)^2} +
(\partial_{r^*}\phi)^2\right) r^2 (1-\mu) d\sigma dr^* dt\\
& \le& (4+) E_{\phi_0}.
\end{eqnarray}

\subsection{The identity, summed}
\label{postermsec}
We introduce the notation
\[
E^{X_{\ge 1}}_{\phi}=
\sum_{\ell\ge 1} E^{X_\ell}_{\phi_{\ell}},
\]
and 
\[
I^{X_{\ge 1}}_{\phi}=
\sum_{\ell\ge 1} I^{X_\ell}_{\phi_{\ell}}.
\]
Our final energy identity for the totality of vector fields $X_\ell$,
$\ell\ge 1$, can be summarized by
the statement
\begin{equation}
\label{Xident}
0\le I^{X_{\ge 1}}_{\phi}(\mathcal{R})=
E^{X_{\ge 1}}_{\phi}(t_1)-E^{X_{\ge 1}}_{\phi}(t_0).
\end{equation}
Defining
\[
I^{X}_{\phi}=\tilde{I}^{X_0}_{\phi_0}+I^{X_{\ge 1}}_{\phi},
\]
\[
E^{X}_{\phi}=(2+)E_{\phi_0}+E^{X_{\ge 1}}_{\phi},
\]
we have the inequality
\[
I^{X}_{\phi}(\mathcal{R})\le |E^{X}_{\phi}(t_1)|+|E^{X}_{\phi}(t_0)|.
\]

\section{The vector field $Y$}
\label{Ysec}
In this section, we shall introduce a vector field which
will give good control on the solution 
near the event horizon $\mathcal{H}^+$.
Although the computations are done in Eddington-Finkelstein coordinates,
the reader is encouraged to compare them with computations in a null coordinate
system which is regular on the event horizon.

We set $Y$ to be the vector
field
\[
Y=-\frac12f\frac{\partial}{\partial u}   -\frac12\tilde{f}\frac{\partial}{\partial v}.
\]
where 
\begin{align*}
f=\frac {\alpha(r^*)}{1-\mu},\qquad
\tilde f = \beta(r^*),
\end{align*}
for functions $\alpha$ and $\beta$ do be defined later.
Below $'$ will denote $\frac{d}{dr^*}$.
We have
\begin{eqnarray*}
T_{\gamma\delta}\pi^{\gamma\delta}&=&-\frac{(\partial_u\phi)^2}{4(1-\mu)^2}
				\left(\frac {\alpha \mu} r - \alpha'\right)
					-\frac{(\partial_v\phi)^2}{4(1-\mu)}\beta'\\
					&&\hbox{}
		-\frac 14|\nabb\phi|^2\left (\frac {\alpha'}{1-\mu} -
		 \frac {\left(\beta(1-\mu)\right)'}{1-\mu}\right)
				+\frac{1}{2r}\left(\frac{\alpha}{1-\mu} -
				\beta\right)\pa_u\phi\, \pa_v\phi.
				\end{eqnarray*}
Integrating in the characteristic 
rectangle $\tilde{\mathcal{R}}'=[u_1,u_2]\times[v_1,v_2]$, we obtain 
\begin{align}
\label{Yident}
\nonumber
&F^Y_\phi (\{u_2\}\times[v_1,v_2])+F^Y_\phi([u_1,\infty]\times \{v_2\})\\
&=
I^Y_\phi(\tilde{\mathcal{R}}')+F^Y_\phi(u_1\times[v_1,v_2])
+F^Y_{\phi}([u_1,\infty]\times\{v_1\})
\end{align}
where 
\begin{eqnarray*}
F^Y_\phi(\{u\}\times[v_1,v_2])\doteq
\int\limits_{v_1}^{v_2}\int\limits_{{\Bbb S}^2} (\alpha|\nabb
\phi|^2+\beta(\partial_v\phi)^2)
(\infty,v) r^2 dv\, d\sigma_{{\mathbb S}^2},
\end{eqnarray*}
\begin{eqnarray*}
F^Y_\phi(\{v\}\times[u_1,u_2])\doteq
\int\limits_{u_1}^{\infty}\int\limits_{{\Bbb S}^2} 
\left (\frac {\alpha}{1-\mu} (\pa_u\phi)^2 + 
 (1-\mu) \beta |\nabb\phi|^2\right) r^2 du\, 
d\sigma_{{\mathbb S}^2},
\end{eqnarray*}
\begin{eqnarray*}
I_\phi^Y(\tilde{\mathcal{R}}')&\doteq&
\int\limits_{v_1}^{v_2} \int\limits_{u_1}^{u_2} \int\limits_{{\Bbb S}^2} 
\left(-\left(\frac{(\partial_u\phi)^2}{(1-\mu)}\left(\frac {\alpha \mu} r - \alpha'\right)
+{(\partial_v\phi)^2}\beta' + 
|\nabb\phi|^2\left ({\alpha'} - {\left(\beta(1-\mu)\right)'}\right)
\right)\right.\\
&&\hbox{}\left.+\frac{2}{r}
\left({\alpha}-\beta (1-\mu)\right)\pa_u\phi\, \pa_v\phi\right)
r^2 du\, dv\, d\sigma_{{\mathbb S}^2}.
\end{eqnarray*}

We define $\alpha$, $\beta$ as follows.
Let $r_0>2M$ be a constant sufficiently close to $2M$,
to be determined by various restrictions that follow.
Set $\alpha=1$ and $\beta=0$ on the event horizon. Furthermore, require
that $\alpha$, $\beta$ 
both be non-negative functions supported in the region $r\le 1.2 r_0$, 
with 
$$
\alpha'(r^*)\sim C (1+|r^*|)^{-1-},\qquad 
\beta(r^*)\sim C (1+|r^*|)^{-1-},\quad \forall r\le r_0
$$
for some constant $C>0$.

We will always apply the above identity in $[u_1,\infty]\times[v_1,v_2]$. 
We have then that
\[
F^Y_\phi(\{\infty\}\times[v_1,v_2])=
\int\limits_{v_1}^{v_2}\int\limits_{{\Bbb S}^2} |\nabb\phi|^2(\infty,v) r^2 dv
d\sigma_{{\mathbb S}^2}.
\]
Let us set 
\[
\tilde{I}^Y_{\phi}(\tilde{\mathcal{R}}')\doteq
\int\limits_{v_1}^{v_2} \int\limits_{u_1}^{\infty} \int\limits_{{\Bbb S}^2} 
\left(\frac{(\partial_u\phi)^2}{(1-\mu)}\left(\frac {\alpha \mu} r - \alpha'\right)
+{(\partial_v\phi)^2}\beta' + |\nabb\phi|^2\left ({\alpha'} - 
{\left(\beta(1-\mu)\right)'}\right)
\right)r^2 du dv d\sigma_{{\mathbb S}^2}.
\]

The quantity $r_0$ will be chosen sufficiently small so that 
all terms in the above integrand for $\tilde{I}^Y_{\phi}$ are nonnegative
in the region $r\le r_0$, and so that moreover
\begin{equation}
\label{cond1ft}
\left(\frac {\alpha \mu} r - \alpha'\right)
\ge  \frac{1}{2r}\left({\alpha}-\beta (1-\mu)\right)^2,
\end{equation}
\begin{equation}
\label{cond2ft}
8r^{-1}(1-\mu)\le \beta'
\end{equation}
in $r\le r_0$. 
In addition, we will require of
$r_0$ that $1.2r_0<3M$ (see Proposition~\ref{giatoY})
and $(\ref{arga})$.

Defining
\[
\hat{I}^Y_{\phi}(\tilde{\mathcal{R}}')\doteq
\int\limits_{v_1}^{v_2} \int\limits_{u_1}^{\infty} \int\limits_{{\Bbb S}^2} 
\frac{2}{r}\left({\alpha}-\beta (1-\mu)\right)\pa_u\phi\, \pa_v\phi\,
r^2 \, du \, dv  \, d\sigma_{{\mathbb S}^2},
\]
we may rewrite $(\ref{Yident})$ as
\begin{align}
\nonumber
\label{Yident2}
&F^Y_\phi (\{\infty\}\times[v_1,v_2])+F^Y_\phi([u_1,\infty]\times \{v_2\})
+\tilde{I}^Y_{\phi}(\tilde{\mathcal{R}}')\\
&=
\hat{I}^Y_\phi(\tilde{\mathcal{R}}')+F^Y_\phi(u_1\times[v_1,v_2])
+F^Y_{\phi}([u_1,\infty]\times\{v_1\}).
\end{align}

In our argument (see Proposition~\ref{giatoY}), the terms $\hat{I}^Y$
and $\tilde{I}^Y$ in the region $r\ge r_0$
will be controlled by $I^X$.
The relation of $F^Y$ with the integrand of $\hat{I}^Y$ (as exploited in
Proposition~\ref{pigeonhole}) 
can be thought of as the manifestation of the red-shift effect as measured
by two local observers.
It is to be compared with the fact that, in equation $(51)$
of~\cite{dr1} for $\pa_v\frac{\zeta}{\nu}$, 
the factor in front of $\frac{\zeta}{\nu}$ on the
right hand side is bounded above and below away from $0$ uniformly in the
region $r\le r_0$.

\section{The vector field $K$}
In this section, we shall define the Morawetz vector field $K$.

We set $K$ to be the vector field 
\[
K= -\frac12u^2\frac{\partial}{\partial u}-\frac12v^2\frac{\partial}{\partial v}.
\]
We have
\[
K^u=-\frac12u^2, K^v=-\frac12v^2, K_u=(1-\mu)v^2, K_v=(1-\mu)u^2.
\]
We compute 
\[
T_{\alpha\beta}\pi^{\alpha\beta}
= t|\nabb\phi|^2\left(\frac12+\frac{\mu r^*}{4r}-\frac{r^*(1-\mu)}{2r}
\right)+ \frac{t r^*(1-\mu)}{4r}\Box \phi^2.
\]
Let $\mathcal{R}=\{t_0\le t\le t_1\}$.
The identity $(\ref{divthe})$ gives 
\[
\hat{I}^K_\phi(\mathcal{R})=\hat{E}^K_{\phi}(t_0)-\hat{E}^K_{\phi}(t_1),
\]
where
\begin{eqnarray*}
\hat{I}^K_\phi(\mathcal{R})&=&
\int_{\mathcal{R}} T_{\alpha\beta}\pi^{\alpha\beta} dVol_{\mathcal{S}}\\
&=&
\int_{t_0}^{t_1}\int_{-\infty}^{\infty}\int_{{\mathbb S}^2}
\left(t|\nabb\phi|^2\left(\frac12+\frac{\mu r^*}{4r}-\frac{r^*(1-\mu)}{2r}
\right)\right.\\
&&\hbox{}
  \left.+ \frac{t r^*(1-\mu)}{4r}\Box \phi^2 \right)\cdot r^2(1-\mu)d\sigma_{{\mathbb S}^2}
   dr^* dt
\end{eqnarray*}
and 
\begin{eqnarray*}
\hat{E}^K_{\phi}(t_i)&=&\int_{\{t=t_i\}} g_{\alpha\beta}n^\alpha P^\beta dVol_{\{t=t_i\}}\\
&=&
\int_{-\infty}^\infty\int_{{\mathbb S}^2}
-\frac1{2\sqrt{1-\mu}}(K^vT_{vu}+K^uT_{uu}+K^uT_{uv}+K^vT_{vv})(r^*,t_i,\omega)\\
&&\hbox{}\cdot r^2\sqrt{1-\mu}\, 
dr^*d\sigma_{{\mathbb S}^2}\\
&=&\int_{-\infty}^\infty\int_{{\mathbb S}^2}\frac1{4\sqrt{1-\mu}}
(u^2T_{uu}+v^2T_{vv}+(u^2+v^2)T_{uv})\\
&&\hbox{}\cdot
r^2\sqrt{1-\mu} \,
dr^*\, d\sigma_{{\mathbb S}^2}\\
&=&\int_{-\infty}^\infty\int_{{\mathbb S}^2}\frac1{4\sqrt{1-\mu}}(u^2(\partial_u\phi)^2+
v^2(\partial_v\phi)^2+(1-\mu)(u^2+v^2)|\nabb \phi|^2)\\
&&\hbox{}\cdot
r^2\sqrt{1-\mu} \,
dr^*\, d\sigma_{{\mathbb S}^2}.
\end{eqnarray*}

In view of the identity
\begin{align*}
\int_\mathcal{R} \psi\Box(\phi^2) dVol_{\mathcal{R}}
&-\int_\mathcal{R} (\Box\psi) \phi^2 dVol_{\mathcal{R}}\\
=&
\int_{\{t=t_1\}} (2g^{\alpha\beta}\phi\nabla_\alpha\phi n_\beta \psi
 -g^{\alpha\beta}\nabla_\alpha\psi n_\beta \phi^2)
 dVol_{\{t=t_1\}}\\
&-
\int_{\{t=t_0\}}(2g^{\alpha\beta}\phi\nabla_\alpha\phi n_\beta \psi
 -g^{\alpha\beta}\nabla_\alpha\psi n_\beta \phi^2) dVol_{\{t=t_0\}},
\end{align*}
applied to
\[
\psi= \frac{tr^*(1-\mu)}{4r},
\]
we obtain 
the identity
\begin{equation}
\label{Kident}
I^K_\phi(\mathcal{R})=E^K_{\phi}(t_1)-E^K_{\phi}(t_0)
\end{equation}
where
\begin{eqnarray}
I^K_\phi(\mathcal{R})=\int_{t_0}^{t_1}\int_{-\infty}^\infty\int_{{\mathbb S}^2} 
&&\left(t|\nabb\phi|^2\left(\frac12+\frac{\mu r^*}{4r}-\frac{r^*(1-\mu)}{2r}
\right)\right.\nonumber\\
&&\hbox{}
  \left. \hskip -7pc+   \frac{t}4\mu r^{-2}|\phi|^2
  \left(2+\frac{r^*(4\mu-3)}r\right) \right)\cdot r^2(1-\mu)\,
dt \, dr^* \, d\sigma_{{\mathbb S}^2}
  \label{eq:IK}
\end{eqnarray}
and
\[
E^K_\phi(t_i)=
\hat{E}^K_\phi(t_i)+
\int_{\{t=t_i\}}\frac{1}{\sqrt{1-\mu}}\left(
\frac{tr^*(1-\mu)}{2r}\phi\partial_t\phi-\phi^2\frac{r^*(1-\mu)}{4r}\right)
\cdot r^2\sqrt{1-\mu}\,
dr^* \, d\sigma_{{\mathbb S}^2}.
\]

For the analogue of the following argument in Minkowski 
space, see \cite{kl}.
Let $S=v\pa_v+u\pa_u$ and $\Sb=v\pa_v-u\pa_u$. Then
$$
v^2 (\pa_v \phi)^2 + u^2(\pa_u \phi)^2= \frac 12 \left ((S\phi)^2 + (\Sb\phi)^2\right),
$$
\begin{align*}
&S=2(t\pa_t + r^*\pa_{r^*}),\qquad \Sb=2 (t\pa_{r^*} + r^*\pa_t ),\\
&2t\pa_t\phi\,\phi= S\phi - 2r^*\pa_{r^*}\phi,\qquad 
2t\pa_t\phi\,\phi = \frac t{r^*} \Sb -2 \frac {t^2}{r^*} \pa_{r^*}\phi.
\end{align*}
Thus, by integration by parts, we 
obtain\footnote{Note that there is no 
problem near the bifurcate sphere in the integration by 
parts argument due to the presence 
of the exponentially decaying $(1-\mu)$ factor.}:
\begin{align*}
\int_{-\infty}^{\infty} 2 t \frac {(1-\mu)r^*}r \pa_t\phi\, \phi r^2 dr^*&= 
\int_{-\infty}^{\infty} \left (r^2 \frac {(1-\mu)r^*}r S\phi\, 
\phi + \pa_{r^*}\left ((1-\mu) r (r^*)^2\right) 
\phi^2\right)dr^*\\&=
\int_{-\infty}^{\infty}  (1-\mu)r^2
\left (\frac {r^*}r S\phi\, 
\phi +\left(\frac {(r^*)^2} {r^2} + 2\frac {r^*}r \right) \phi^2\right)dr^*.
\end{align*}
It now follows that 
\begin{align*}
\int_{-\infty}^{\infty}\int_{{\mathbb S}^2}& \left (\frac 14 (1-\mu) (S\phi)^2+ 
2 t \frac {(1-\mu)r^*}r \pa_t\phi\, \phi 
 -\frac {(1-\mu)r^*}r \phi^2\right) r^2\, dr^* d\sigma_{{\mathbb S}^2}\\ 
 &\qquad =\int_{-\infty}^{\infty}  (1-\mu) r^2 \left (\left (\frac 12 S\phi + 
 \frac {r^*}r\phi\right)^2 +
\frac {(r^*)^2}{r^2} \phi^2\right) \,  dr^* d\sigma_{{\mathbb S}^2}.
\end{align*}
We may thus write
\begin{eqnarray}
\label{UCyIldIz}
\nonumber
E^K_\phi(t_i)&=&
\frac14\int_{-\infty}^{\infty}\int_{{\mathbb S}^2}\frac{1}{\sqrt{1-\mu}}
\left(\frac14(1+\mu)(S\phi)^2
+\frac12(\Sb\phi)^2+(1-\mu)(u^2+v^2)|\nabb\phi|^2\right.\\
&&\hbox{}+\left.(1-\mu)\left(\frac12S\phi+\frac
{r^*}r\phi\right)^2+(1-\mu)\frac{(r^*)^2}{r^2}\phi^2\right)\cdot r^2\sqrt{1-\mu}
dr^*\, d\sigma_{{\mathbb S}^2}.
\end{eqnarray}
In particular, $E^K_{\phi}(t_i)$ is nonnegative.

\section{Comparison estimates}
Let us introduce one final spacetime integral quantity:
For $\mathcal{X}$ a region denote
\[
I_{\phi}(\mathcal{X})=
\int_{{\mathcal X}} \left (|\partial_{r^*}\phi|^2+|\partial_t\phi|^2+(1-\mu)|
\nabb\phi|^2\right ) dVol.
\]

We have the following
\begin{proposition}
\label{firstp}
If $\mathcal{X}$ is a rectangle defined by
\[
\mathcal{X}=\{t_0\le t\le t_1\}\cap \{2M<\hat{r}_0\le r\le \hat{R}\},
\]
then 
\begin{equation}\label{eq:loss}
I_\phi(\mathcal{X})\le C(I^X_\phi(\mathcal{X})+I^X_{\phi_\omega}(\mathcal{X})).
\end{equation}
Moreover, for either $\hat{R}<3M$ or $\hat{r}_0>3M$,
 \begin{equation}\label{eq:noloss}
I_\phi(\mathcal{X})\le CI^X_\phi(\mathcal{X}).
\end{equation}
\end{proposition}
\begin{proof}
From the definition of 
the functions $f_\ell$, we have that for $\ell\ge1$,
\begin{equation}
\label{ct1}
\int_{\mathcal{X}}
\frac{\phi_{\ell}^2}{(1-\mu) \left (1+(r^*)^2\right)^2}  
dVol \le
C I^X_{\phi_{\ell}}(\mathcal{X}),
\end{equation}
and thus, summing over $\ell\ge 1$, we obtain
\begin{equation}
\label{ct1a}
\int_{\mathcal{X}}\frac{\phi_{\ge 1}^2}{(1-\mu) 
\left (1+(r^*)^2\right)^2}  
dVol \le
C I^X_{\phi_{\ge1}}(\mathcal{X}).
\end{equation}
Now using this bound, we can apply $(\ref{XLdef})$ again with the function
\begin{equation}\label{eq:f}
f(r^*)=\int_{0}^{r^*}\frac{d\rho}{1+(\rho)^2}.
\end{equation}
Note that this choice ensures the non-negativity of the space-time 
integral containing the angular derivatives. 
We obtain 
\begin{equation}
\label{ct3pre}
\int_{\mathcal{X}} 
\frac {|\partial_{r^*}\phi_{\ge 1}|^2}
{(1-\mu)\left (1+(r^*)^2\right)}   dVol + \int_{\mathcal{X}}   \frac { 
f(r^*)\,(2-3\mu)|\nabb\phi_{\ge 1}|^2}
{2r}   dVol \le CI^X_{\phi_{\ge 1}}(\mathcal{X}).
\end{equation}
Applying $(\ref{nova2})$, using the representation 
\begin{equation}\label{eq:repr}
\Box \left (\phi_{\ge 1}^2\right ) = 
\frac 2{1-\mu} \left (-(\pa_t\phi_{\ge 1})^2 + 
(\pa_{r^*}\phi_{\ge 1})^2 + 
(1-\mu)|\nabb\phi_{\ge 1}|^2\right)
\end{equation}
and the function 
$\tilde{f}(r^*)=f(r^*)\cdot (2-3\mu)(2r)^{-1}$,
we obtain  from 
 $(\ref{ct1a})$ and  $(\ref{ct3pre})$  the bound 
\begin{equation}
\label{cti}
 \int_{\mathcal{X}}     f(r^*)\,(2-3\mu)(2r)^{-1}|\pa_t \phi_{\ge 1}|^2
  dVol \le CI^X_{\phi_{\ge 1}}(\mathcal{X}).
\end{equation} 

On the other hand, by $(\ref{thedefi})$, we clearly have
\begin{equation}
\label{ct3pre2}
\int_{\mathcal{X}}  
\frac {|\partial_{r^*}\phi_{0}|^2\, \mu^2}
{(1-\mu)\left (1+(r^*)^2\right)^{1+}}   dVol  + 
\int_{\mathcal{X}}   \frac {|\partial_t\phi_{0}|^2\, \mu^2}
{(1-\mu)\left (1+(r^*)^2\right)^{1+}}  
 dVol \le C\tilde{I}^X_{\phi_{0}}(
\mathcal{X}).
\end{equation}
Estimates $(\ref{ct3pre})$, $(\ref{cti})$ and
$(\ref{ct3pre2})$ immediately yield
\begin{align}
\label{ct3}
\nonumber
\int_{\mathcal{X}} &  {\frac{ f(r^*) (2-3\mu)
|\partial_{t}\phi|^2\, \mu^2}{2r(1+(r^*)^2)^{1+}}} dVol 
+ 
\int_{\mathcal{X}}   
\frac {|\partial_{r^*}\phi|^2\, \mu^2}
{(1-\mu)\left (1+(r^*)^2\right)^{1+}}   dVol\\
& + \int_{\mathcal{X}}   \frac { 
f(r^*)\,(2-3\mu)|\nabb\phi|^2}
{2r}   dVol\le CI^X_{\phi}.
\end{align}
This clearly gives $(\ref{eq:noloss})$. 

Applying our argument again to the solution $\phi_\omega=\Omega\phi$ 
of the wave equation
$\Box\phi_\omega=0$, where $\Omega$ is an angular momentum
operator, in view of the 
fact that this solution has vanishing zeroth spherical
harmonic, we obtain as in $(\ref{ct1a})$
\begin{equation}
 \label{ct2}
\int_{\mathcal{X}}\frac{r^2 |\nabb\phi|^2}{(1-\mu) \left 
(1+(r^*)^2\right)^2}  dVol \le
C I^X_{\phi_\omega}(\mathcal{X}).
\end{equation}
Here $I_{\phi_\omega}$ actually denotes the sum of $I$ applied
to $\Omega\phi$ where $\Omega$ range over an appropriate basis
of angular momentum operators.
Then, using \eqref{eq:repr} and \eqref{ct3pre2}, and arguing 
as before we also obtain that 
\begin{equation}\label{eq:cit}
\int\frac{|\pa_t \phi|^2\, \mu^2 }{(1-\mu)\left (1+(r^*)^2\right)^2}  
dVol \le
C \left (I^X_\phi(\mathcal{X})+ I^X_{\phi_\omega}(\mathcal{X})\right ).
\end{equation}
The estimates $(\ref{ct3})$, $(\ref{ct2})$ and $(\ref{eq:cit})$ together give \eqref{eq:loss}.
\end{proof}
\begin{proposition}
\label{withhardy}
Consider the hypersurface $\{t=t_i\}$. We have
\[
|E^X_{\phi}(t_i)|\le C E_{\phi}.
\]
\end{proposition}
\begin{proof}
In view of the Cauchy-Schwarz inequality applied to
the boundary terms defined by $(\ref{Xbtdef})$,
it suffices to 
to establish the Hardy inequality
\begin{equation}
\label{Hardy}
\int\limits_{-\infty}^\infty \int\limits_{{\Bbb S}^2} (1+|r^*|)^{-2} |\phi|^2 r^2 dr^* 
d\sigma_{{\mathbb S}^2}
\le C \int\limits_{-\infty}^\infty \int\limits_{{\Bbb S}^2} |\pa_{r^*} \phi|^2 r^2 dr^* 
d\sigma_{{\mathbb S}^2}.
\end{equation}

Define the function $f(r^*)$ by solving the equation 
$$
f'=\left (\frac {r}{1+|r^*|}\right)^2
$$
with the boundary condition $f(-\infty)=0$, i.e., 
$$
f(r^*)=\int\limits_{-\infty}^{r^*} \left (\frac {r(\rho)}{1+|\rho|}\right)^2 d\rho.
$$
Here $'$ denotes $\frac{d}{dr^*}$.
Clearly, 
$$
f(r^*)\sim |r^*|^{-1},\quad ({\text{for}}\,\, r^*\to 
-\infty)\qquad{\text{and}}\qquad
f(r^*)\sim r^*,\quad ({\text{for}}\,\, r^*\to +\infty).
$$
Now,
\begin{eqnarray*}
\int\limits_{-\infty}^\infty\int_{{\mathbb S}^2} (1+|r^*|)^{-2} |\phi|^2 r^2 dr^* &=&
\int\limits_{-\infty}^\infty\int_{{\mathbb S}^2} f'(r^*) |\phi|^2  dr^* \\
&=& 
-2 \int\limits_{-\infty}^\infty\int_{{\mathbb S}^2}
 f(r^*) \pa_{r^*} \phi\, \phi\,  dr^*\\ &\le& 
2  \left (\,\int\limits_{-\infty}^\infty\int_{{\mathbb S}^2} 
\frac {\left (f(r^*)\right)^2}{f'(r^*)} 
|\pa_{r^*}\phi|^2  dr^* \right)^{\frac 12}\\
&&\hbox{}\cdot
\left (\,\int\limits_{-\infty}^\infty\int_{{\mathbb S}^2} 
f'(r^*) |\phi|^2  dr^* \right)^{\frac 12}\\ &=&
2  \left (\,\int\limits_{-\infty}^\infty\int_{{\mathbb S}^2} 
\frac {\left ((1+|r^*|) f(r^*)\right)^2}{r^2} 
|\pa_{r^*}\phi|^2  dr^* \right)^{\frac 12}\\
&&\hbox{}\cdot
\left (\,\int\limits_{-\infty}^\infty\int_{{\mathbb S}^2}
 (1+|r^*|)^{-2} |\phi|^2 r^2 dr^* \right)^{\frac 12}.
\end{eqnarray*}
Inequality $(\ref{Hardy})$ follows immediately from
$$
 \frac {\left ((1+|r^*|) f(r^*)\right)^2}{r^2} \le C r^2.
$$
\end{proof}
We will denote by $E_{\phi_\omega}$ the total energy associated 
with the solutions of the wave equation $\phi_\omega=\Omega\phi$,
obtained by applying a basis angular momentum operators $\Omega$ to 
$\phi$, and summing. 
We will also set $E_{\phi,\phi_\omega}=E_\phi+E_{\phi_\omega}$ etc.

\begin{proposition}
\label{bndX1}
Consider the rectangle $\mathcal{X}$ of Proposition~\ref{firstp}. We have
\[
I_\phi(\mathcal{X})\le CE_{\phi,\phi_\omega}.
\]
If $\mathcal{X}\cap \left ([t_0,t_1]\times \{r=3M\}\right )=0$ then 
\[
I_\phi(\mathcal{X})\le CE_{\phi}.
\]
\end{proposition}
\begin{proof}
This follows immediately from the previous two Propositions.
\end{proof}

\begin{proposition}
\label{metosxnma}
Consider the rectangle $\tilde{\mathcal{R}}=[u_1,\infty]\times[v_1,v_2]$, and
the shaded triangle $\mathcal{R}$ depicted. 
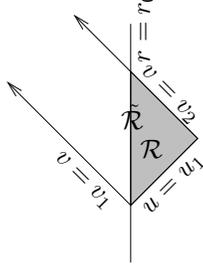
\[
\input{duoperioxes.pstex_t}
\]
We have 
\[
F_{\phi}^Y(\{u_1\}\times[v_1,v_2])\le C(\varpi_\phi(u_1,v_2)-\varpi_\phi(u_1,v_1)),
\]
\[
\varpi_\phi(\infty,v_2)-\varpi_\phi(\hat{u}(v_2),v_2)\le
CF_{\phi}^Y([u_1,\infty)\times \{v_2\}),
\]
\[
\int\limits_{\hat u(v_2)}^\infty  \int\limits_{{\Bbb S}^2}
\frac {(\pa_u\phi)^2}{1-\mu} d\sigma_{{\Bbb S}^2}  du\le CF_{\phi}^Y([u_1,\infty)\times \{v_2\})
\]
\[
 \int\limits_{v_1}^{\hat v(u)} \int\limits_{{\Bbb S}^2} 
|\nabb\phi|^2 d\sigma_{{\Bbb S}^2} dv \le F_{\phi}^Y(\{u\}\times[v_1,\hat v(u)]),
\quad \forall u\ge u_1
\]
where $\hat{u}(v_2)$ and $\hat v(u)$ are defined so that $r(\hat{u}(v_2),v_2)=r_0$ and 
$r(u,\hat v(u))=r_0$ respectively.
\end{proposition}
\begin{proof}
This follows immediately from the definition of $Y$ in Section \ref{Ysec}. 
\end{proof}

The next 
proposition shows that in the identity $(\ref{Yident2})$ generated by 
vector field $Y$, the  
space-time terms without a sign can be controlled with the help of 
$I^X$.
\begin{proposition}
\label{giatoY}
With $\tilde{\mathcal{R}}$, $\mathcal{R}$ as above, we have
\begin{align*}
&F_\phi^Y(\{\infty\}\times\{v_1,v_2\})+F_\phi^Y([u_1,\infty)\times\{v_2\})
+\frac12\tilde{I}^Y_{\phi}(\tilde{\mathcal{R}})\\
&\le C \left(I^X_{\phi}(\mathcal{R})+F_{\phi}^Y([u_1,\infty]\times\{v_1\})
+\varpi_\phi(u_1,v_2)-\varpi_\phi(u_1,v_1)\right).
\end{align*}
\end{proposition}
\begin{proof}
By Proposition~\ref{metosxnma}, we have
that 
\[
F_{\phi}^Y(\{u_1\}\times[v_1,v_2])
\le C(\varpi_\phi(u_1,v_2)-\varpi_\phi(u_1,v_1)).
\]
From $(\ref{Yident2})$, it would suffice then to show that
\begin{equation}
\label{itsufficesto}
\hat{I}^Y_{\phi}(\tilde{\mathcal{R}})\le CI^X_{\phi}(\mathcal{R})
+\frac12\tilde{I}^Y_{\phi}(\tilde{\mathcal{R}}).
\end{equation}

Let us decompose: 
\[
\hat{I}^Y_\phi(\tilde{\mathcal{R}})=\hat{I}^Y_\phi(\tilde{\mathcal{R}}\setminus
\mathcal{R})+ \hat{I}^Y_\phi(\mathcal{R}).
\]
By an application of Cauchy-Schwarz and $(\ref{cond1ft})$, 
$(\ref{cond2ft})$, 
we have in the region $\tilde{\mathcal{R}}\setminus
\mathcal{R}$ that
\begin{eqnarray*}
\frac{2}r(\alpha-\beta(1-\mu))\partial_u\phi\partial_v\phi
&\le& 
\frac12\left(
\frac{(\alpha-\beta(1-\mu))^2}{2r(1-\mu)}(\partial_u\phi)^2
+8r^{-1}(1-\mu)(\partial_v\phi)^2\right)\\
&\le&\frac12\left(\frac{(\partial_u\phi)^2}{(1-\mu)}\left(\frac {\alpha \mu} r - \alpha'\right)
+\beta'(\partial_v\phi)^2\right),
\end{eqnarray*}
so integrating, we obtain
\[
\hat{I}^Y_\phi(\tilde{\mathcal{R}}\setminus
\mathcal{R})\le \frac12 \tilde{I}^Y_\phi(\tilde{\mathcal{R}}\setminus
\mathcal{R}).
\]
On the other hand,  by Proposition~\ref{firstp} and 
the condition $1.2r_0<3M$, 
we easily see that
\[
\hat{I}^Y_{\phi}(\mathcal{R})\le C(I^X_{\phi}(\mathcal{R})).
\]
Since 
\[
0\le \tilde{I}^Y_\phi(\tilde{\mathcal{R}}\setminus
\mathcal{R})=\tilde{I}^Y_\phi(\tilde{\mathcal{R}}')-
\tilde{I}^Y_\phi(\mathcal{R}),
\]
and 
\[
|\tilde{I}^Y_\phi(\mathcal{R})|\le C(I^X_{\phi}(\mathcal{R})),
\]
$(\ref{itsufficesto})$ follows immediately.
\end{proof}

The next proposition
can be thought of as a pigeonhole argument,
which will allow us to pick
values of $v$ where the boundary terms generated by the vector field $Y$ 
gain additional decay.
\begin{proposition}
\label{pigeonhole}
With $\tilde{\mathcal{R}}$ as above, we have
\[
\inf_{v_1\le v\le v_2}F^Y_\phi([u_1,\infty]\times\{v\})
\le C(v_2-v_1)^{-1}\tilde{I}^Y_{\phi}(\tilde{\mathcal{R}})
+C(\varpi_\phi(u_1,v)-\varpi_\phi(\hat{u}(v),v)).
\]
\end{proposition}
\begin{proof}
This follows immediately from the fact that $\int_{a}^b f\ge (b-a)\inf f$,
and the inequality
\begin{eqnarray*}
F^Y_\phi([u_1,\infty]\times\{v\})&\le& C\int_{u_1}^{\infty}\int_{{\mathbb S}^2}
\left(\frac{(\partial_u\phi)^2}{(1-\mu)}\left(\frac {\alpha \mu} r - \alpha'\right)\right.
\\
&&\hbox{}+\left.
{(\partial_v\phi)^2}\beta' + |\nabb\phi|^2\left ({\alpha'} - {\left(\beta(1-\mu)\right)'}\right)
\right)r^2 du d\sigma\\
&&\hbox{}+\varpi_\phi(u_1,v)-\varpi_\phi(\hat{u}(v),v).
\end{eqnarray*}
\end{proof}
The next proposition
estimates the positive part of  $I^K$
(i.e.~the terms with the wrong sign) by 
$I(\mathcal{X})$, with a certain loss:
\begin{proposition}
\label{Kcomparison}
Let $\tilde{\mathcal{R}}'$ and $\mathcal{X}$ denote regions
$\tilde{\mathcal{R}}'=\{ t_0\le t\le t_1\}$,
$\mathcal{X}=\tilde{\mathcal{R}}'\cap\{ r_0\le r\le R\}$ depicted below:
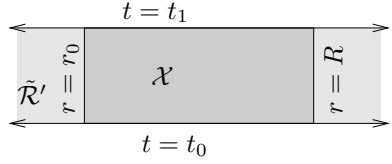
\[
\input{trapezoeides2.pstex_t}
\]
where $R$ is sufficiently large.
We have
\[
I^K_{\phi}(\tilde{\mathcal{R}}')\le t_1\, I_{\phi}(\mathcal{X}).
\]
\end{proposition}
\begin{proof}
Recall that 
\begin{eqnarray*}
I^K_\phi(\tilde {\mathcal R}')&=&
\int\limits_{t_0}^{t_1}\int\limits_{-\infty}^\infty \int\limits_{{\Bbb 
S}^2} 
 \left (\frac 
t2 |\nabla_\omega \phi|^2  \left (1+\frac{3\mu-2}{2r} r^*\right)\right.\\
&&\hbox{}\left.+\frac 
t4\left ( \frac {2\mu}{r^2} + \frac {(4\mu-3)\mu}{r^3}r^*\right )\phi^2 
\right)r^2 (1-\mu)\, dt\, dr^*\, d\sigma_{{\Bbb S}^2}.
\end{eqnarray*}
We have required of $r_0$ that 
\begin{equation}
\label{arga}
r_0^*<\max\left(-\frac{2r_0}{4\mu_0-3}, -\frac{2r_0}{3\mu_0-2}\right)<0.
\end{equation}
We require of $R$ that $R$ is sufficiently large, to
be determined below, and, in particular, that $R>r_2$ where
$r_2$ is the infimum of all $r$ satisfing, for all $\bar{r}\ge r$,
the inequality
\begin{equation}
\label{arga2}
\bar{r}^*>\min\left( -\frac{2\bar{r}}{4 \bar\mu-3}, -\frac{2\bar r 
}{3\bar \mu-2}
\right) >0.
\end{equation}
Note that $r_2<infty$.
The choices $(\ref{arga})$, $(\ref{arga2})$ ensure that 
in the regions $\{r\le r_0\}$, $\{r\ge R\}$, the integrand
of $I_\phi^K$ is nonpositive. Thus, 
we have
\begin{eqnarray*}
I_\phi^K(\tilde{\mathcal{R}}')&=&I_\phi^K(\tilde{\mathcal{R}}'\setminus\mathcal{X})
			+I_\phi^K(\mathcal{X})\\
			&\le&I_\phi^K(\mathcal{X}).
\end{eqnarray*}
On the other hand, one sees easily that
\[
\int_{{\mathcal X}} \frac 
t2 |\nabla_\omega \phi|^2  \left (1+\frac{3\mu-2}{2r} r_*\right)\, dVol \le  C\, t_1 I_\phi(\mathcal{X}).
\]
It will then suffice to prove the inequality 
\[
\int_{{\mathcal X}} \frac t4\left ( \frac {2\mu}{r^2} + \frac {(4\mu-3)\mu}{r^3}r_*\right )\phi^2\, dVol
\le -\epsilon I_\phi^K(\tilde{\mathcal{R}}'\setminus\mathcal{X}) + C\, t_1 I_\phi(\mathcal{X}).
\]
for some sufficiently small $\epsilon>0$.
The function
$$
H(r^*)=\left ( \frac {2\mu}{r^2} + \frac {(4\mu-3)\mu}{r^3}r^*\right )
$$
is positive only  in a subset of
the region of $2M<r_0<r_1\le r\le r_2<R$.
Moreover, $H(r^*)$ behaves like 
$r^{-3}$ as $r\to +\infty$.
The desired result will then follow from the following inequality:
$$
\int\limits_{r_1^*}^{r_2^*} \int\limits_{{\Bbb S}^2}  \phi^2 r^2(1-\mu) d\sigma_{{\Bbb S}^2} dr^* \le 
C \int\limits_{r_1^*}^{r_3^*} \int\limits_{{\Bbb S}^2}  (\phi')^2 r^2(1-\mu) d\sigma_{{\Bbb S}^2} dr^* +
c \int\limits_{r_2^*}^{r_3^*} \int\limits_{{\Bbb S}^2} 
 r^{-3} \phi^2\,  r^2(1-\mu) d\sigma_{{\Bbb S}^2} dr^*
$$
for some small constant $c>0$ and some large $r_3\le R$. 
This in turn easily follows from rescaling
the following one dimensional 
estimate:
\begin{lemma}
For any $c>0$ there exist sufficiently large constants $C$ and $A$ such that
$$
\int_{1}^2 \phi^2(x) dx 
\le C\int_1^A (\phi')^2 dx + 2 c \int_1^A x^{-1} \phi^2 dx.
$$
\end{lemma}
\begin{proof}
We write 
$$
\phi^2(x)-\phi^2(y)= 2\int_x^y \phi' \phi\, dz\le 2 \left (\int_x^y z\, \phi'^2 dz\right)^{\frac 12} 
 \left (\int_x^y z^{-1} \phi^2\, dz\right)^{\frac 12} .
$$
 Thus
 $$
 \int_{1}^2 \phi^2(x) dx \le c^{-2} \int_1^y z\, \phi'^2 dz + c^2 \int_1^y z^{-1} \phi^2\, dz
 + \phi^2(y).
 $$
 Dividing by $y$ and integrating in $y$ in 
the region $[2,A]$ with $A=e^{1/c}$, we obtain
 $$
(c^{-1}-\log 2) \int_{1}^2 \phi^2(x) dx \le c^{-3} 
\int_1^{A}  z\, \phi'^2 dz + c \int_1^{A} z^{-1} \phi^2\, dz
 + \int_{2}^{A} y^{-1} \phi^2(y)\, dy.
 $$
The desired conclusion follows immediately with $C\sim Ac^{-2}$. 
\end{proof}
\end{proof}
\begin{proposition}
\label{Kcomparison2}
For the region $\tilde{\mathcal{R}}'$ as above, we have
\[
E^K_{\phi}(t_1)\le 
C(I^K_{\phi}(\tilde{\mathcal{R}}')+ E^K_{\phi}(t_0)).
\]
\end{proposition}
\begin{proof}
This follows immediately from $(\ref{Kident})$.
\end{proof}

\begin{proposition}
On a constant $t_1$ hypersurface, we have
the bounds
\begin{align}
\label{Kbd0}
\nonumber
&\int_{-\infty}^\infty\int_{{\mathbb S}^2} 
\frac{1}{\sqrt{1-\mu}}\left(u^2(\partial_u\phi)^2 
+v^2(\partial_v\phi)^2\right.\\
&\qquad+\left.(1-\mu)(u^2+v^2)|\nabb\phi|^2\right)
\cdot r^2 \sqrt{1-\mu}\, dr^* d\sigma_{{\mathbb S}^2}\le E^K_{\phi}(t_1),
\end{align}
\begin{equation}
\label{Kbd1}
\int_{-\infty}^\infty\int_{{\mathbb S}^2} \sqrt{1-\mu} 
\frac{r_*^2}{r^2}\phi^2 \cdot r^2 \sqrt{1-\mu}\, dr^* d\sigma_{{\mathbb S}^2}
\le E^K_\phi(t_1),
\end{equation}
\begin{equation}
\label{Kbd2}
\int_{-\infty}^\infty\int_{{\mathbb S}^2} \sqrt{1-\mu}
\frac{t^2}{r^2}\phi^2 \cdot r^2\sqrt{1-\mu}\, dr^* d\sigma_{{\mathbb S}^2}
\le CE^K_\phi(t_1).
\end{equation}
\end{proposition}
\begin{proof}
The bounds $(\ref{Kbd0})$ and $(\ref{Kbd1})$  
follow immediately from $(\ref{UCyIldIz})$.

For $(\ref{Kbd2})$, write
\begin{align*}
\int 2 t \frac {(1-\mu)r^*}r \pa_t\phi\, \phi r^2 dr^*&= 
\int \left (r^2 \frac {(1-\mu)t}r \Sb\phi\, \phi + t^2 
\pa_{r^*}\left ((1-\mu) r\right) \phi^2\right)\\&=
\int  (1-\mu)r^2
\left (\frac {t}r \Sb\phi\, \phi +\frac {t^2} {r^2}\phi^2\right)\\ &=
\int  (1-\mu)r^2\left (\frac 12 \Sb\phi +\frac tr \phi\right)^2 - 
\frac 14 \int  (1-\mu)r^2 (\Sb\phi)^2,
\end{align*}
and use \eqref{UCyIldIz}.
\end{proof}

We now compare the boundary terms $E_\phi^K(t_i)$
 generated by $K$ with the flux of $\frac{\pa}{\pa t}$-energy.
\begin{proposition}
With $\tilde{\mathcal{R}}'$ as before, let $(r^*_1,t_1)$, $(\tilde{r}^*_1, t_1)$,
be so as for $t_1-\tilde{r}^*_1\ge 1$, $t_1+r^*_1\ge 1$. Then
\label{t^-2gain}
\[
\varpi_\phi(\tilde r^*_1,t_1)-\varpi_\phi(r^*_1,t_1)\le (t_1-\tilde{r}^*_1)^{-2}E^K_{\phi}(t_1)
						+(t_1+r^*_1)^{-2}E^K_{\phi}(t_1).
\]
\end{proposition}
\begin{proof}
This follows immediately from $(\ref{Kbd0})$ and the geometry of the region considered.
\end{proof}
\begin{proposition}
\label{giatoX}
With $\mathcal{X}$ as above, suppose 
\begin{equation}
\label{choices}
t_1\le 1.1t_0, \qquad |r^*(r_0)|+|r^*(R)|\le .1t_0.
\end{equation}
Then we have
\begin{eqnarray*}
I^X_{\phi}(\mathcal{X})&\le& 
Ct_0^{-2}E^K_\phi(t_0).
\end{eqnarray*}
\end{proposition}
\begin{proof}
Let $\chi$ be a smooth cut-off function equal to one on the interval 
$[-1,1]$ and vanishing outside the interval $[-1.5,1.5]$.
Define $\psi$ to be a solution of the wave equation $\Box_g \psi=0$ with 
initial conditions 
$$
\psi(t_0,r^*) = \chi(2r^*/t_0) \phi(t_0,r^*),\quad 
\pa_t \psi(t_0,r_*) = \chi(2r^*/t_0) \pa_t \phi(t_0,r^*).
$$
We 
claim that 
$$
I^X_\phi({\cal X})=I^X_\psi({\cal X}).
$$
This follows from the choices $(\ref{choices})$ and the fact that initial data
for $\psi(t_0,\cdot)$ coincides with those 
of $\phi(t_0,\cdot)$ for the values of $-.5 t_0\le r^*\le .5 t_0$.
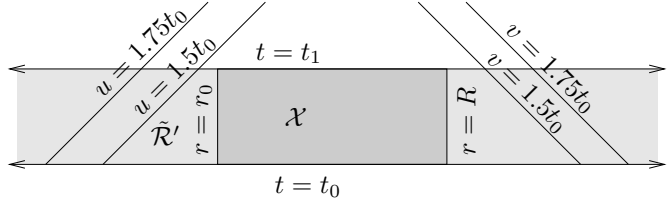
\[
\input{poincare.pstex_t}
\]

We now apply Proposition~\ref{bndX1} to the function $\psi$ to obtain 
$$
I^X_\psi(\tilde{\cal R}')\le C E_\psi(t_0).
$$
It remains to show that 
$$
E_\psi(t_0)\le C t_0^{-2} E_\phi^K(t_0).
$$

From the definition of the energy $E_\psi$ and the properties of the cut-off
function $\chi$ we see immediately that
$$
E_\psi(t_0)\le \varpi_\phi(-.75 t_0,t_0)-\varpi_\phi(.75t_0,t_0) + 
C t_0^{-2}\int\limits_{-.75 t_0}^{.75 t_0}\int\limits_{{\Bbb S}^2} 
|\phi|^2 r^2 d\sigma \, dr^*.
$$
In view of Proposition~\ref{t^-2gain}, it suffices to show that 
$$
\int\limits_{-.75 t_0}^{.75 t_0}\int\limits_{{\Bbb S}^2} |\phi|^2 r^2 d\sigma \, dr^*\le 
C E_\phi^K(t_0).
$$
We will rely on the one-dimensional inequality 
$$
\int\limits_{-a}^a |f(x)|^2 dx \le C a^2 \left (\, \int\limits_{-a}^a |\pa_x f(x)|^2 dx  +
\int\limits_{-1}^1 |f(x)|^2 dx \right). 
$$
Applying this to the function $r \phi$, and then integrating over ${\mathbb S}^2$,
we obtain
\begin{eqnarray*}
\int\limits_{-.75 t_0}^{.75 t_0}\int\limits_{{\Bbb S}^2} |\phi|^2 r^2 d\sigma \, dr^*
&\le& 
C t_0^2 \left ( \int\limits_{-.75 t_0}^{.75 t_0}\int\limits_{{\Bbb S}^2}
 \left (|\pa_{r_*}\phi|^2 + \frac {(1-\mu)^2}{r^2} |\phi|^2\right ) r^2 d\sigma \, dr^*
 \right.\\
&&\hbox{}\left.
+ \int\limits_{-1}^{1}\int\limits_{{\Bbb S}^2} |\phi|^2 r^2 d\sigma \, dr^*\right)\\
&\le&CE_{\phi}^K(t_0),
\end{eqnarray*}
where the last inequality follows from $(\ref{Kbd0})$--$(\ref{Kbd2})$.
\end{proof}

\section{Local observers' uniform energy boundedness}
For any $v_2\ge1$ construct a rectangle $\mathcal{\tilde{R}}$ 
and triangle $\mathcal{R}$ as in the
figure of Proposition~\ref{metosxnma} with $v_1=1$.
Let 
$$
H_\phi = \int\limits_{-\infty}^{\infty}\int\limits_{{\mathbb S}^2}
r^2(1-\mu)^{-\frac12}\left((\partial_t\phi)^2+(\partial_{r^*}\phi)^2+
(1-\mu) |\nabb\phi|^2\right)
(1,r^*,\sigma_{{\Bbb S}^2}) dr^* 
d\sigma_{{\mathbb S}^2}.
$$

Similar to $E_{\phi_\omega}, E_{\phi,\phi_\omega}$ etc. we define 
$H_{\phi_\omega}$, $H_{\phi,\phi_\omega}$ etc. 
Note that $H_{\phi,\phi_\omega,\phi_{\omega\omega\omega}}\le 
\bar{E}_0$.
By Cauchy stability\footnote{To see this, one must actually change 
coordinates.}, we have that 
\[
F^Y_{\phi}([u_1,\infty]\times\{1\})\le
CH_{\phi}.
\]
On the other hand, by Proposition~\ref{bndX1}, 
 we have the uniform estimate
\[
I^X_{\phi}(\mathcal{R})\le CE_{\phi}
\]
and thus, by
Proposition~\ref{giatoY}, we have
\begin{equation}
\label{gre0}
\tilde{I}^Y_{\phi}(\mathcal{\tilde{R}})\le C\left(E_{\phi}+H_{\phi} \right),
\end{equation}
\begin{equation}
\label{gre}
F^Y_\phi([u_1,\infty]\times\{v_2\})\le C\left(E_{\phi}+H_{\phi}\right),
\end{equation}
\[
F_{\phi}^Y(\{\infty\}\times[1,v_2])\le C(E_{\phi}+H_{\phi}).
\]
Note: Using the above bounds, the classical
result of Kay and Wald can be reproven without
exploiting discrete isometries of Schwarzschild.
See Section \ref{Kay}.

\section{Proof of Theorem~\ref{cvmt}}
Set $t_0=1$, $t_{i+1}=1.1t_i$, and let $u_i$, $v_i$ be defined
by the relations $r(u_i,v_i)=r_0$, $u_i+v_i=t_i$. 
Define the sets
\[
\tilde{\mathcal{R}}_i=[u_i,\infty]\times[v_i,v_{i+1}], 
\]
\[
\tilde{\mathcal{R}}'_i=\{t_i\le t\le t_{i+1}\},
\]
\[
\tilde{\mathcal{S}}_i'=\cup_{j=0}^i\tilde{\mathcal{R}}'_i,
\]
\[
\mathcal{X}_i=\tilde{\mathcal{R}}'_i\cap\{r_0\le r\le R\},
\]
\[
\mathcal{Y}_i=\cup_{j=0}^i\mathcal{X}_i.
\]
\subsection{Energy decay}
Applying first Proposition~\ref{bndX1}  to $\tilde{\mathcal{Y}}_i$,
we obtain
\[
I_\phi^X(\tilde{\mathcal{Y}}_i) \le CE_\phi,
\]
\[
I_{\phi_{\omega}}^X(\tilde{\mathcal{Y}}_i)\le CE_{\phi_{\omega}}.
\]
Applying now Propositions~\ref{Kcomparison} and~\ref{Kcomparison2}
with $\tilde{\mathcal{R}}'=\tilde{\mathcal{S}}_i'$, $\mathcal{X}=\mathcal{Y}_i$,
we obtain
\[
E_\phi^K(t_i)\le Ct_iE_{\phi,\phi_\omega}+CE_\phi^K(1),
\]
and similarly
\[
E_{\phi_\omega}^K(t_i)\le Ct_iE_{\phi_\omega,\phi_{\omega\omega}}+
			CE_{\phi_\omega}^K(1).
\]

Applying now Proposition~\ref{giatoX} to $\mathcal{X}_i$, we obtain
\[
I^X_\phi(\mathcal{X}_i)\le Ct_i^{-1}E_{\phi,\phi_\omega}
				+Ct_i^{-2}E_\phi^K(1),
\]
and also
\[
I^X_{\phi_\omega}(\mathcal{X}_i)\le Ct_i^{-1}E_{\phi_{\omega},\phi_{\omega\omega}}
				+Ct_i^{-2}E_{\phi_{\omega}}^K(1).
\]
Applying Proposition~\ref{firstp} and once again Proposition~\ref{Kcomparison}, but now with
 $\tilde{\mathcal{R}}'=\tilde{\mathcal{R}}'_i$,
 $\mathcal{X}=\mathcal{X}_i$,
we obtain
\[
I_{\phi}^K(\tilde{\mathcal{R}}'_i)\le CE_{\phi,\phi_\omega,\phi_{\omega\omega}}
				+Ct_i^{-1}E_{\phi,\phi_\omega}^K(1).
\]
Summing over $i$, one obtains
\[
I_\phi^K(\mathcal{Y}_i)=\sum_{j=0}^i I_\phi^K(\mathcal{X}_i)
\le 
C(\log t_i) 
E_{\phi,\phi_\omega,\phi_{\omega\omega}}+CE_{\phi,\phi_\omega}^K(1).
\]
Proposition~\ref{Kcomparison2} then gives
\[
E_{\phi}^K(t_i)\le C(\log t_i) E_{\phi,\phi_\omega,\phi_{\omega\omega}}+
			CE_{\phi,\phi_\omega}^K(1),
\]		
and similarly,
\[
E_{\phi_\omega}^K(t_i)\le C(\log t_i) E_{\phi_{\omega},\phi_{\omega\omega},
			\phi_{\omega\omega\omega}}+
			CE_{\phi_{\omega},\phi_{\omega\omega}}^K(1).
\]
One repeats the procedure one final time to remove the $\log$ term:
\begin{equation}
\label{finalK}
E_{\phi}^K(t_i)\le C (E_{\phi,\phi_\omega,\phi_{\omega\omega},\phi_{\omega\omega\omega}}+
			E_{\phi,\phi_\omega,\phi_{\omega\omega}}^K(1)) = C \bar E_1.
\end{equation}
Note one also obtains
then from Proposition~\ref{giatoX}
that
\begin{equation}
\label{kiautoxreiazetai}
I_\phi^X(\mathcal{X}_i)\le C \bar E_1 t_i^{-2}.
\end{equation}
Proposition~\ref{t^-2gain}, $(\ref{finalK})$ and energy conservation 
now immediately imply $(\ref{Cfb})$.

\subsection{Local observers' energy decay}
Refer to the following diagramme:
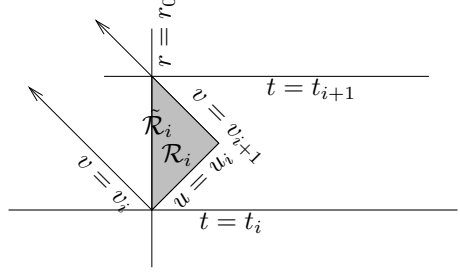
\[
\input{decay.pstex_t}
\]
From $(\ref{gre0})$, 
we have already shown 
\[
\tilde{I}^Y_\phi(\mathcal{R}_i)\le C (E_\phi+ H_\phi).
\]
Apply Proposition~\ref{pigeonhole} together with $(\ref{Cfb})$ to obtain
the existence of a $\tilde{v}_i$ such that
\begin{equation}
\label{choice}
F_\phi^Y([u_1,\infty]\times\{v\})\le C 
(E_\phi+H_\phi)(v_{i+1}-v_{i})^{-1}\le C (E_\phi+H_\phi) t_i^{-1}.
\end{equation}
Now let $\tilde{u}_i$ be defined by $r(\tilde{u}_i,\tilde{v}_i)=r_0$,
and construct rectangle $\tilde{\hat{\mathcal{R}}}_i$ and triangle
$\hat{\mathcal{R}}_i$, as depicted: 
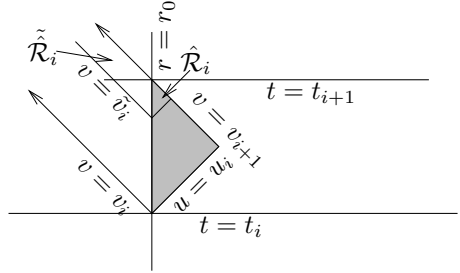
\[
\input{decay2.pstex_t}
\]
By Proposition~\ref{metosxnma} and $(\ref{choice})$, 
we have that 
\[
\varpi_\phi(\tilde{u}_i,\tilde{v}_i)-\varpi_\phi(\infty,\tilde{v}_i)
\le F_\phi^Y([u_1,\infty]\times\{\tilde{v}_i\})\le C (E_\phi+H_\phi) 
t_i^{-1}.
\]
On the other hand, by
$(\ref{Cfb})$ and the fact that $u\sim v\sim t_i$ on 
$\tilde{\mathcal{R}}_i$, we have 
\[
\varpi_\phi(\tilde{u}_i,v_{i+1})-\varpi_\phi(\tilde{u}_{i+1},\tilde{v}_i))
\le C \bar E_2 t_i^{-2}.
\]
Finally, from $(\ref{kiautoxreiazetai})$, we have the bound
\[
I^X_\phi(\hat{\mathcal{R}}_i)\le  C \bar E_1 t_i^{-2}.
\]
Thus, we obtain from Proposition~\ref{giatoY} and the inequality
$E_\phi+H_\phi\le \bar{E}_0$, the bound
\[
F_\phi^Y([u_i,\infty]\times\{v_{i+1}\})\le C( \bar E_1+\bar E_0) 
t_{i}^{-1}.
\]
We now repeat the procedure to obtain
\[
F_\phi^Y([u_i,\infty]\times\{v_i\})\le C( \bar E_1+\bar E_0) t_{i}^{-2}.
\]

Further application of Proposition~\ref{giatoY} now yields
\begin{equation}
\label{otherdir}
F_\phi^Y([u,\infty]\times\{v\})\le  C( \bar E_1+\bar E_0) v^{-2},
\end{equation}
\begin{equation}
\label{goodforangular}
F_\phi^Y(\{u\}\times[v,\tilde{v}])\le C( \bar E_1+\bar E_0) v^{-2}.
\end{equation}
Note that, by Proposition \ref{metosxnma}, 
this implies in particular, that if $(u,\tilde{v})$ is such that
$r(u,\tilde{v})\le r_0$ and if $(\tilde u,v)$ is such that 
$r(\tilde u, v)\le r_0$
then
\begin{align}
&\int_v^{\tilde{v}}\int_{{\mathbb S}^2}|\nabb\phi|^2d\sigma_{{\Bbb S}^2} dv\le 
C( \bar E_1+\bar E_0)v^{-2},\label{inpartic}\\
&\int_{\tilde u}^{\infty}\int_{{\mathbb S}^2}
\frac {(\pa_u \phi)^2}{1-\mu} d\sigma_{{\Bbb S}^2} du\le 
C( \bar E_1+\bar E_0) v^{-2}.
\label{eq:unav}
\end{align}
Recall that these $L^2$ estimates are not available in this region
from the usual energy estimate.

More generally one easily shows
the following
\begin{theorem}
Let $\mathcal{S}$ be an achronal hypersurface in the closure of the
exterior. Then 
$F_\phi^Y(\mathcal{S})\le C( \bar E_1+\bar E_0) 
(v_+(\mathcal{S}))^{-2}$, 
where
$v_+(\mathcal{S})$ is as in the statement of Theorem~\ref{main}, and 
$F_\phi^Y(\mathcal{S})$ denotes the flux of $T_{\alpha\beta}Y^\alpha$ 
through $\mathcal{S}$.
\end{theorem}
The above theorem applies in particular to subsets of the event horizon $\mathcal{H}^+$.

\subsection{Uniform boundedness of $\phi$}
\label{Kay}
We now reprove the classical result of Kay and Wald stated as estimate 
\eqref{Cub} of Theorem \ref{cvmt}.
First consider the region $r\ge r_0$. We have
$$
|\phi(t,r^*,\omega)|\le \int_{r^*}^\infty |\pa_{r^*} 
\phi(t,\rho,\omega)|\,d\rho
\le r^{-\frac 12} 
\left (\int_{r^*}^\infty |\pa_{r^*} \phi(t,\rho,\omega)|^2\,r^2(\rho)\, 
d\rho\right )^{\frac 12}.
$$
By the Sobolev embedding on ${\Bbb S}^2$,
\begin{align*}
\int\limits_{r^*}^\infty |\pa_{r^*} \phi(t,\rho,\omega)|^2\,r^2(\rho)\, 
d\rho \le& C
\int\limits_{r^*}^\infty \left (|\pa_{r^*} \phi(t,\rho,\omega)|^2+
|\pa_{r^*} \Omega \phi(t,\rho,\omega)|^2\right.\\
& \left.+
|\pa_{r^*} \Omega^2 \phi(t,\rho,\omega)|^2\right) 
r^2(\rho)\,d\sigma_{{\Bbb S}^2} d\rho
\\ \le& C E_{\phi,\phi_\omega,\phi_{\omega\omega}}.
\end{align*}
Thus, for $r\ge r_0$
\begin{equation}\label{eq:less}
|\phi|^2\le C r^{-1} E_{\phi,\phi_\omega,\phi_{\omega\omega}}.
\end{equation}
In the region 
$r\le r_0$ we define $\tilde u(v)$ so that $r(\tilde u(v),v)=r_0$. Then 
\begin{eqnarray*}
|\phi(u,v,\omega)|&\le& |\phi(\tilde u(v),v,\omega)+ 
\int_{\tilde u(v)}^u |\pa_u\phi(u',v,\omega)|\, du'\\
&\le&
|\phi(\tilde u(v),v,\omega)|+c \left (\int_{\tilde u(v)}^\infty \frac 
{(\pa_u\phi(u',v,\omega))^2}{1-\mu}\, 
du'\right)^{\frac 12}.
\end{eqnarray*}
Since $r(\tilde u(v),v)=r_0$, by \eqref{eq:less} we have
$$
|\phi(\tilde u(v),v,\omega)|^2\le C r_0^{-1} 
E_{\phi,\phi_\omega,\phi_{\omega\omega}}.
$$
On the other hand, by the Sobolev embedding, we have 
\begin{align*}
\int_{\tilde u(v)}^\infty \frac {(\pa_u\phi(u',v,\omega))^2}{1-\mu}\, du'\
\le& C \int_{\tilde u(v)}^\infty \frac 1{1-\mu}
\left ({(\pa_u\phi(u',v,\omega))^2}+
 {(\pa_u \Omega\phi(u',v,\omega))^2}\right.\\
&\left.
+{(\pa_u\Omega^2 \phi(u',v,\omega))^2}\right) d\sigma_{{\Bbb S}^2} du'\\ 
\le& 
C\left (E_{\phi,\phi_\omega,\phi_{\omega\omega}}+ H_{\phi,\phi_\omega,\phi_{\omega\omega}}\right),
\end{align*}
where the last inequality follows from Proposition \ref{metosxnma} and \eqref{gre}.
Thus, for $r\le r_0$,
$$
|\phi|^2\le C\left (E_{\phi,\phi_\omega,\phi_{\omega\omega}}+ H_{\phi,\phi_\omega,\phi_{\omega\omega}}\right).
$$
Combining the estimates for $r\ge r_0$ and $r\le r_0$ and observing that 
$\bar E_0=E_{\phi,\phi_\omega,\phi_{\omega\omega}}+ H_{\phi,\phi_\omega,\phi_{\omega\omega}}$
we obtain the desired result.

\subsection{Pointwise decay for $\phi$}

\subsubsection{$|\phi|\le Cv_+^{-1}$ decay near $\mathcal{H}^+$}
For a fixed large $\hat{R}$ consider the region 
\[
\mathcal{U}=\{(u,v)\in \{r\le 
\hat{R}\}:|\phi(\tilde{u},\tilde{v},\omega)|<
\hat{C}\tilde{v}_+^{-1}, \forall\tilde{u},\,\, 
\tilde{v}\le v, \,\, r(\tilde{u},\tilde{v})\le \hat{R}\},
\]
for a $\hat{C}$ to be chosen later. 
$\mathcal{U}$ is clearly open. Moreover, for sufficiently
large $\hat{C}$, $\mathcal{U}\supset\{r\le \hat{R}\}\cap\{v\le 2\}$.
We will show that $\mathcal{U}$ is closed, and thus,
coincides with $r\le \hat{R}$. 
For this, clearly it suffices to show
\begin{equation}
\label{suffices}
|\phi|\le (\hat{C}/2)v_+^{-1}
\end{equation}
in $\mathcal{U}\cap \{v\ge 2\}$.

Let $(\tilde u,\tilde v)\in\mathcal{U}\cap\{v\ge 2\}$, and consider
the null segment $\{\tilde u\}\times[\tilde v-1, \tilde v]$. Note that
this null segment is contained in $r\le \hat{R}$.
We have a Sobolev inequality
\begin{eqnarray*}
|\phi|^2(u,v,\omega) &\le& K\, \int\limits_{\tilde{v}-1}^{\tilde{v}}
\int\limits_{{\Bbb S}^2} \left (|\phi|^2+|\pa_v\phi|^2+
|\nabb\phi|^2+ |\nabb\pa_v \phi|^2 \right.\\
&&\hbox{}+
\left.|\nabb\nabb\phi|^2+|\nabb\nabb\pa_v\phi|^2 \right) 
d\sigma_{{\Bbb S}^2}\, dv,
\end{eqnarray*}
in $r\le \hat{R}$.
Our energy decay estimate $(\ref{Cfb})$ and our local observers' energy decay
estimate $(\ref{goodforangular})$, in particular $(\ref{inpartic})$,
imply that, after commuting the equation twice 
with angular momentum operators $\Omega$,
\[
\int\limits_{\tilde{v}-1}^{\tilde{v}}\int\limits_{{\Bbb S}^2} \left 
(|\pa_v\phi|^2+
|\nabb\phi|^2+ |\nabb\pa_v\phi|^2 +
|\nabb\nabb\phi|^2+|\nabb\nabb\pa_v\phi|^2 \right)
d\sigma_{{\mathbb S}^2}\, dv \le  \bar E_2 \tilde{v}^{-2},
\]
in $r\le \hat{R}$.
Choosing $\hat{C}$ sufficiently large, it remains to show, say that 
\[
 \int\limits_{\tilde{v}-1}^{\tilde{v}}\int\limits_{{\Bbb S}^2} |\phi|^2 
(\tilde{u},v,\omega)
 d\sigma_{{\mathbb S}^2}\, dv \le 
\left(\frac {\hat{C}^2}{8K}\right)\tilde{v}^{-2}.
\]

Let $u_0({v})$ be defined by $r^*(u_0(v),v)=v-\tilde v+ \hat{R}^*+1$, in 
particular
$u_0(v)+v=\tilde v-\hat{R}^*-1=t$ is independent of $v$.
We have
\begin{align*}
\int\limits_{\tilde{v}-1}^{\tilde{v}}
\int\limits_{{\Bbb S}^2}& |\phi|^2(u,v,\omega) d\sigma_{{\mathbb S}^2}\,dv 
= 
\int\limits_{\tilde{v}-1}^{\tilde{v}}
\int\limits_{{\Bbb S}^2} |\phi|^2(u_0,v,\omega) d\sigma_{{\mathbb 
S}^2}\,dv + 2 \int\limits_{v_1}^{v_2}
\int\limits_{u_0}^u \int\limits_{{\Bbb S}^2}
\pa_u \phi\,\phi\, d\sigma_{{\mathbb S}^2} du'\,dv\\ &\le 
\int\limits_{\tilde{v}-1}^{\tilde{v}}\int\limits_{{\Bbb S}^2} |\phi|^2(u_0,v) d\sigma_{{\Bbb S}^2}\,dv +  
\int\limits_{\tilde{v}-1}^{\tilde{v}}\int\limits_{u_0}^\infty \int\limits_{{\Bbb S}^2}
\frac{(\pa_u \phi)^2}{1-\mu} d\sigma_{{\mathbb S}^2}\, du'\, dv\\
&\qquad+ 
\int\limits_{\tilde{v}-1}^{\tilde{v}}\int\limits_{u_0}^\infty \int\limits_{{\Bbb S}^2}
(1-\mu) \phi^2\, d\sigma_{{\mathbb S}^2} du'\, dv.
\end{align*}
From the bounds
$(\ref{Kbd2})$, $(\ref{finalK})$, we have
\[
\int\limits_{\tilde{v}-1}^{\tilde{v}}
\int\limits_{{\Bbb S}^2} |\phi|^2(u_0(v),v) d\sigma_{{\Bbb S}^2}\,dv\le  
\bar CE_1 \tilde{v}^{-2}.
\]
On the other hand, from our local observers' energy estimate \eqref{eq:unav}, we have 
$$
\int\limits_{u_0}^\infty \int\limits_{{\Bbb S}^2}
\frac{(\pa_u \phi)^2}{1-\mu} d\sigma_{{\Bbb S}^2}\, du\le C
( \bar E_1+\bar E_0) \tilde{v}^{-2}.
$$
Thus, choosing $\hat{C}$ sufficiently large,
we are left with showing, say, the bound 
\begin{equation}
\label{leftwith}
\int\limits_{\tilde{v}-1}^{\tilde{v}}\int\limits_{u_0(v)}^\infty \int\limits_{{\Bbb S}^2}
(1-\mu) \phi^2\, d\sigma_{{\Bbb S}^2} du'\, dv\le 
\left(\frac{\hat{C}^2}{16 K}
\right) \tilde{v}^{-2}.
\end{equation}

We have the inclusion
\[
[u_0(v),\infty)\times[\tilde{v}-1,\tilde{v}]
\subset
\{(t,r^*):\, \tilde{v}-\hat{R}^*-1\le t<\infty,\quad \tilde{v}-t-1\le 
r^*\le 
\tilde{v}-t\}.
\]
Thus,
$$
\int\limits_{\tilde v-1}^{\tilde v}\int\limits_{u_0(v)}^\infty \int\limits_{{\Bbb S}^2}
(1-\mu) |\phi|^2\, d\sigma_{{\mathbb S}^2} du'\, dv\le C 
\int\limits_{\tilde{v}-\hat{R}^*-1}^\infty\int\limits_{\tilde{v}-t-1}^{\tilde{v}-t} 
\int\limits_{{\Bbb S}^2}
(1-\mu) |\phi|^2 d\sigma_{{\mathbb S}^2}\, dr^*\, dt.
$$
We may write
\begin{eqnarray*}
\int\limits_{\tilde{v}-\hat{R}^*-1}^\infty
\int\limits_{\tilde{v}-t-1}^{\tilde{v}-t} \int\limits_{{\Bbb S}^2}
(1-\mu) |\phi|^2 d\sigma_{{\mathbb S}^2}\, dr^*\, dt 
&=& \int\limits_{\tilde{v}-\hat{R}^*-1}^{\tilde{v}-\hat{R}^*-1+A}
\int\limits_{\tilde{v}-t-1}^{\tilde{v}-t} \int\limits_{{\Bbb S}^2}
(1-\mu) |\phi|^2 d\sigma_{{\mathbb S}^2}\, dr^*\, dt \\
&&\hbox{}+
\int\limits_{\tilde{v}-\hat{R}^*-1+A}^\infty\int\limits_{\tilde{v}-t-1}
^{\tilde{v}-t} 
\int\limits_{{\Bbb S}^2}
(1-\mu) |\phi|^2 d\sigma_{{\mathbb S}^2}\, dr^*\, dt .
\end{eqnarray*}
The first integral can be estimated with the help of $(\ref{Kbd2})$, $(\ref{finalK})$, 
\begin{eqnarray}
\nonumber
\label{there}
 \int\limits_{\tilde{v}-\hat{R}^*-1}^{\tilde{v}-\hat{R}^*-1+A}
\int\limits_{\tilde{v}-t-1}^{\tilde{v}-t} \int\limits_{{\Bbb S}^2}
(1-\mu) |\phi|^2 d\sigma_{{\Bbb S}^2}\, dr^*\, dt &\le& A(1+ 
|\tilde{v}-\hat{R}^*-1|)^{-2}
C\sup_{t\in [\tilde v-\hat{R}^*-1, \tilde v-\hat{R}^*-1+A]} E^K_\phi(t) \\
&\le&  AC\bar E_1 (\hat{R}^*)^2\, \tilde{v}^{-2}.
\end{eqnarray}
For the second integral we use the fact that, for $A$ large enough,
the region of integration above is contained in $\mathcal{U}$, as 
$r^*(u,v)\le \hat{R}^*-A+1$ and $\tilde v-1\le v\le \tilde v$ there,
 so we
have
\[
|\phi|\le \hat{C}v^{-1}.
\]
On the other hand, we have $(1-\mu)\le h e^{r^*}$ in this region.
Thus
$$
\int\limits_{\tilde{v}-\hat{R}^*-1+ 
A}^\infty\int\limits_{\tilde{v}-t-1}^{\tilde{v}-t} 
\int\limits_{{\Bbb S}^2}
(1-\mu) |\phi|^2 d\sigma_{{\mathbb S}^2}
\, dr^*\, dt \le\hat{C}^2  \tilde{v}^{-2} 
\int\limits_{-\infty}^{-A+\hat{R}^*+1}
e^{r^*}\, dr^*\le h\hat{C}^2 e^{-A+\hat{R}^*+1} \tilde{v}^{-2}.
$$
Choosing $A$ to be sufficiently large, say 
$A=10+\log h+\hat{R}^*$, and $\hat{C}$ large
as before, and satisfying in addition, say 
$$
\hat{C}\ge 32AK (\hat{R}^*)^2 \left (\bar E_2 + \bar E_1 + 
\bar{E}_0\right),
$$
 we obtain $(\ref{leftwith})$, as desired. Note that  $\bar E_2$ dominates 
 $ \bar E_1$ and $\bar E_0$.

\subsubsection{Decay in $r\ge \hat{R}$}
We turn to $r\ge \hat{R}$.
First consider the region $\{r\ge \hat{R}\}\cap\{u\ge 1\}$. By 
the Sobolev inequality,
$$
r^2 |\phi(u,v,\omega)|^2\le C \int\limits_{{\Bbb S}^2} |\phi|^2 r^2 d\sigma_{{\mathbb S}^2} + 
C \int\limits_{{\Bbb S}^2}
 \left( |r \nabb \phi|^2 + |r^2 \nabb\nabb \phi|^2\right)
 r^2 d\sigma_{{\mathbb S}^2}.
 $$
For any $\hat{R}\le \tilde r\le \hat{R}+1$, for $k=0,1,2$,
and any $r$ we have 
 \begin{align*}
 \int\limits_{{\Bbb S}^2}
 |\Omega^k \phi|^2 r^2(t,r,\omega)  d\sigma_{{\mathbb S}^2} \le& 
\int\limits_{{\Bbb S}^2}
 |\Omega^k \phi|^2  r^2(t,\tilde{r}^*,\omega) d\sigma_{{\mathbb S}^2} \\
&+ C
 \int\limits_{\tilde{r}^*}^{r^*}\int\limits_{{\Bbb S}^2}
 \left (|\partial_{r^*} \Omega^k \phi| \, |\Omega^k\phi|+ r^{-1} 
 |\Omega^k \phi|^2\right ) r^2\,
 d\rho \,d\sigma_{{\mathbb S}^2} .
 \end{align*}
Applying a pigeonhole argument in $\tilde r$ and 
$(\ref{Kbd2})$, $(\ref{finalK})$ with
$\Omega^k\phi$, we obtain
that $\tilde{r}$ can be chosen so as for
$$
 \int\limits_{{\Bbb S}^2}
 |\Omega^k \phi|^2 r^2(t, \tilde{r}^*,\omega)
d\sigma_{{\Bbb S}^2} \le 
\bar E_2 \, t^{-2}.
$$
On the other hand,
 from $(\ref{Kbd0})$, $(\ref{Kbd2})$, and $(\ref{finalK})$ applied
to $\Omega^k\phi$,
 we have 
\[
\int\limits_{r_0^*}^{r^*}\int\limits_{{\Bbb S}^2}|\Omega^k \phi|^2 r\,
 d\rho\,d\sigma_{{\mathbb S}^2} \le \bar E_2 r\, t^{-2},
\]
\[
\int\limits_{r_0^*}^{r^*}\int\limits_{{\Bbb S}^2}
|\partial_{r^*} \Omega^k \phi| \, |\Omega^k\phi|\, r^2\,
 d\rho \,d\sigma_{{\mathbb S}^2}\le \bar E_2 \, r \, t^{-1} u^{-1} .
\]
 Therefore, since in the region $u\ge 1$ we have that (say) $t\ge r/2$ 
, it follows that
 $$
 |\phi|\le \bar E_2 (r\,t\,u)^{-\frac 12}\le 2 \bar E_2 r^{-1}u^{-\frac12}.
 $$
 On the other hand, since $\max (r,u)\ge v/2$ we also obtain
 $$
 |\phi|\le \bar E_2 (1+|v|)^{-1}.
 $$

 The region $\{r\ge r_0\}\cap \{u\le 1\}\cap\{t\ge 1\}$ can be treated 
 similarly by integrating out to spatial 
 infinity using the estimates
 \[
 \int_{r^*}^{\infty}\int_{{\mathbb S}^2}
 |\partial_{r^*} \Omega^k \phi| \, |\Omega^k\phi|\, r^2\,
 d\rho \,d\sigma_{{\mathbb S}^2}\le \bar E_2 \, |u|^{-1}, 
 \]
 \[
  \int_{r^*}^{\infty}\int_{{\mathbb S}^2}|\Omega^k \phi|^2 r\,
 d\rho\,d\sigma_{{\mathbb S}^2} \le \bar E_2 \, r^{-1}.
 \]
In this region, one obtains 
 $$
 |\phi|\le C\, r^{-1} (1+|u|)^{-\frac 12}, 
 \qquad  |\phi|\le \bar E_2 \, v^{-1} (1+|u|)^{-\frac 12}.
 $$
This completes the proof.

\section{Acknowledgements}
MD thanks Princeton University and IR thanks the
University of Cambridge for their hospitality during various visits
when this research was carried out. The authors both
thank the Newton Institute, where this research was
completed during the programme ``Global Problems 
in Mathematical Relativity''. MD is supported in part by
DMS-0302748. IR is supported in part by DMS-0406627.

\end{document}